\documentstyle[11pt,epsfig,dina4p]{article} 

\begin{document}
\headheight 12pt
\renewcommand{\floatpagefraction}{0.1}
\renewcommand{\textfraction}{0.1}
\def\rZ{{\rm Z}}
\def\rW{{\rm W}}
\def\rG{{\rm GUT}}
\def\rS{{\rm SUSY}}
\def\rH{{\rm Higgs}}
\def\rF{{\rm Fam}}
\def\MG{M_\rG}
\newcommand{\bsg}  {{\ifmmode b\rightarrow s\gamma
                     \else $b\rightarrow s\gamma$ \fi}}
\newcommand{\Bbsg}  {{\ifmmode BR(\b\rightarrow s\gamma)
\else $BR(b\rightarrow s\gamma)$ \fi}}
\newcommand{\smdchi}    {smdchi.eps}                  
\newcommand{\figochiall}{chi2_25_n.eps}               
\newcommand{\rb}[1]{\raisebox{1.5ex}[-1.5ex]{#1}}
\newcommand{\smas}[2]{\tilde{m}^#2_{#1}}
\newcommand{\figI}      {mttb}                
\newcommand{\figII}     {m1_m2}                 
\newcommand{\figIII}    {chi2_gyebdl}           
\newcommand{\figIV}     {spectrum}              
\newcommand{\figV}      {bsg}                   
\newcommand{\figVI}     {relic}                 
\newcommand{\figVII}    {mchar1}                
\newcommand{\figVIII}   {chi2_gyel}             
\newcommand{\figIX}     {dmb}                   
\newcommand{\figX}      {mt_tb1}                
\newcommand{\figXa}     {running_At}
\newcommand{\figXI}     {mglui}                 
\newcommand{\figXII}    {mu_mglu}               
\newcommand{\figXIII}   {atop_mglu}             
\newcommand{\figXIV}    {abot_mglu}             
\newcommand{\figXV}     {atau_mglu}             
\newcommand{\figXVI}    {mu}                    
\newcommand{\figXVII}   {mh}                    
\newcommand{\figXVIII}  {ma}                    
\newcommand{\figXIXa}   {m1}                    
\newcommand{\figXIXb}   {m2}                    
\newcommand{\figXIXc}   {S1_m1}                 
\newcommand{\figXIXd}   {S2_m2}                 
\newcommand{\figXIXe}   {dmz}                   
\newcommand{\figXX}     {43a_hz0}               
\newcommand{\figXXI}    {49a_hz0-300}           
\newcommand{\figXXII}   {x_hz0_low_high}        
\newcommand{\figXXIII}  {49a_chichi-300}        
\newcommand{\mc}{Monte Carlo }
\newcommand{\mcs}{Monte Carlos }
\newcommand{\brem}{brems\-strah\-lung }
\newcommand{\bq}{\begin{equation}}
\newcommand{\eq}{\end{equation}}
\newcommand{\ba}{\begin{array}}
\newcommand{\ea}{\end{array}}
\newcommand{\bqa}{\begin{eqnarray}}
\newcommand{\eqa}{\end{eqnarray}}
\newcommand{\nn}{\nonumber \\}
\newcommand{\mpmm}{\mu^{+}\mu^{-}}
\newcommand{\tptm}{\tau^{+}\tau^{-}}
\newcommand{\sq}{^{2}}
\newcommand{\etal}{\it et al.\rm}
\newcommand{\ra}{\rightarrow}
\newcommand{\lnf}{{\ifmmode \Lambda^{(N_f)} \else $\Lambda^{(N_f)}$\fi}}
\newcommand{\ms}{{\ifmmode \overline{MS} \else $\overline{MS}$\fi}}
\newcommand{\dr}{{\ifmmode \overline{DR} \else $\overline{DR}$\fi}}
\newcommand{\lms}{{\ifmmode \Lambda^{(5)}_{\overline{MS}}
                 \else $\Lambda^{(5)}_{\overline{MS}}$\fi}}
\newcommand{\lam}{{\ifmmode \Lambda \else $\Lambda$\fi}}
\newcommand{\gev}{{\ifmmode {\rm GeV} \else ${\rm GeV}$\fi}}
\newcommand{\gevc}{{\ifmmode {\rm GeV/c^2} \else ${\rm GeV/c^2}$\fi}}
\newcommand{\tev}{{\ifmmode {\rm TeV} \else ${\rm TeV}$\fi}}
\newcommand{\tevc}{{\ifmmode {\rm TeV/c^2} \else ${\rm TeV/c^2}$\fi}}
\newcommand{\lp}{{\ifmmode L^+  \else $L^+$\fi}}
\newcommand{\lm}{{\ifmmode L^-  \else $L^-$\fi}}
\newcommand{\mlp}{{\ifmmode M(L^-)  \else $M(L^-)$\fi}}
\newcommand{\mlz}{{\ifmmode M(L^0)  \else $M(L^0)$\fi}}
\newcommand{\lz}{{\ifmmode L^0     \else $L^0$\fi}}
\newcommand{\ev}{{\ifmmode GeV/c^2       else $GeV/c^2$\fi}}
\newcommand{\tri}{{\ifmmode \triangleup  \else $\triangleup$\fi}}
\newcommand{\unl}{{\ifmmode U_{lL^0}  \else $U_{lL^0}$\fi}}
\newcommand{\gL}{{\ifmmode g_L  \else $g_{L}$\fi}}
\newcommand{\gR}{{\ifmmode g_R  \else $g_{R}$\fi}}
\newcommand{\gumu}{{\ifmmode \gamma^{\mu}  \else $\gamma^{\mu}$\fi}}
\newcommand{\gunu}{{\ifmmode \gamma^{\nu}  \else $\gamma^{\nu}$\fi}}
\newcommand{\gdmu}{{\ifmmode \gamma_{\mu}  \else $\gamma_{\mu}$\fi}}
\newcommand{\gdnu}{{\ifmmode \gamma_{\nu}  \else $\gamma_{\nu}$\fi}}
\newcommand{\stw}{{\ifmmode\sin^2\theta_W  \else $\sin^{2}\theta_{W}$
\fi}}
\newcommand{\sws}{{\ifmmode \;\sin^2\theta_W  \else $\;\sin^{2}\theta_{W}$
\fi}}
\newcommand{\cws}{{\ifmmode \;\cos^2\theta_W  \else $\;\cos^{2}\theta_{W}$
\fi}}
\newcommand{\sw}{{\ifmmode \;\sin\theta_W  \else $\sin\theta_{W}$
\fi}}
\newcommand{\cw}{{\ifmmode \;\cos\theta_W  \else $\;\cos\theta_{W}$
\fi}}
\newcommand{\tw}{{\ifmmode \;\tan\theta_W  \else $\;\tan\theta_{W}$
\fi}}
\newcommand{\qq}{{\ifmmode q\overline{q} \else $q\overline{q}$\fi}}
\newcommand{\lR}{{\ifmmode l_R  \else $l_R$\fi}}
\newcommand{\lL}{{\ifmmode l_L  \else &l_L$\fi}}
\newcommand{\nt}{{\ifmmode \nu_{\tau} \else $\nu_{\tau}$\fi}}
\newcommand{\nuR}{{\ifmmode \nu_R  \else $\nu_R$\fi}}
\newcommand{\nuL}{{\ifmmode \nu_L  \else $\nu_L$\fi}}
\newcommand{\qR}{{\ifmmode g_R  \else $q_R$\fi}}
\newcommand{\qL}{{\ifmmode q_L  \else $q_L$\fi}}
\newcommand{\qRp}{{\ifmmode q_R'  \else $q_{R}$'\fi}}
\newcommand{\qLp}{{\ifmmode q_L'  \else $q_{L}$'\fi}}
\newcommand{\est}{{\ifmmode e^{\bf \ast}  \else $e^{\bf \ast}$\fi}}
\newcommand{\lst}{{\ifmmode l^{\bf \ast}  \else $l^{\bf \ast}$\fi}}
\newcommand{\must}{{\ifmmode \mu^{\bf \ast}  \else $\mu^{\bf \ast}$\fi}}
\newcommand{\taust}{{\ifmmode \tau^{\bf \ast}  \else $\tau^{\bf \ast}$
\fi}}
\newcommand{\pperp}{{\ifmmode p_t  \else $p_t$\fi}}
\newcommand{\et}{{\ifmmode E_t  \else $E_t$\fi}}
\newcommand{\xt}{{\ifmmode x_t  \else $x_t$\fi}}
\newcommand{\smumu}{{\ifmmode \sigma_{\mu\mu}  \else $\sigma_{\mu\mu}$
\fi}}
\newcommand{\eg}{{\ifmmode e\gamma  \else $e\gamma$\fi}}
\newcommand{\epem}{{\ifmmode e^+e^-  \else $e^+e^-$\fi}}
\newcommand{\lplm}{{\ifmmode L^+L^-  \else $L^+L^-$\fi}}
\newcommand{\pp}{{\ifmmode p\overline p  \else $p\overline p$\fi}}
\newcommand{\llz}{{\ifmmode L^0\overline{L}^0 \else
$L^0\overline{L}^0$\fi}}
\newcommand{\epemt}{{\ifmmode e^+e^- \to  \else $e^+e^- \to$\fi}}
\newcommand{\eb}{{\ifmmode E_{beam}  \else $E_{beam}$\fi}}
\newcommand{\ip}{{\ifmmode pb^{-1}  \else $pb^{-1}$\fi}}
\newcommand{\upm}{{\ifmmode ^{\pm}  \else $^{\pm}$\fi}}
\newcommand{\de}{{\ifmmode ^{\circ}  \else $^{\circ}$ \fi}}
\newcommand{\appr}{{\ifmmode \sim \else $\sim$ \fi}}
\newcommand{\corresp}{{\ifmmode \stackrel{\wedge}{=}
                      \else   $\stackrel{\wedge}{=}$ \fi}}
\newcommand{\sqrts}{{\ifmmode \sqrt{s} \else $\sqrt{s}$\fi}}
\newcommand{\zz}{{\ifmmode Z^0  \else $Z^0$\fi}}
\newcommand{\mz}{{\ifmmode M_{Z}  \else $M_{Z}$\fi}}
\newcommand{\mzs}{{\ifmmode M_{Z}^2  \else $M_{Z}^2$\fi}}
\newcommand{\mw}{{\ifmmode M_{W}  \else $M_{W}$\fi}}
\newcommand{\mws}{{\ifmmode M_{W}^2  \else $M_{W}^2$\fi}}
\newcommand{\mh}{{\ifmmode M_{Higgs}  \else $M_{Higgs}$\fi}}
\newcommand{\gt}{{\ifmmode \Gamma_{tot} \else $\Gamma_{tot}$\fi}}
\newcommand{\msusy}{{\ifmmode M_{SUSY}  \else $M_{SUSY}$\fi}}
\newcommand{\msusys}{{\ifmmode M_{SUSY}^2  \else $M_{SUSY}^2$\fi}}
\newcommand{\su}{{\ifmmode SU(3)_C\otimes\- SU(2)_L\otimes\- U(1)_Y  \else $SU(3)_C\otimes SU(2)_L\otimes U(1)_Y$\fi}}
\newcommand{\suthree}{{\ifmmode SU(3)_C  \else $SU(3)_C$\fi}}
\newcommand{\sutwo}{{\ifmmode  SU(2)_L\otimes U(1)_Y \else $SU(2)_L\otimes U(1)_Y$\fi}}
\newcommand{\taup} {{\ifmmode \tau_{proton} \else $\tau_{proton}$\fi}}
\newcommand{\agut}{{\ifmmode \alpha_{GUT}  \else $\alpha_{GUT}$\fi}}
\newcommand{\mgut}{{\ifmmode M_{GUT}  \else $M_{GUT}$\fi}}
\newcommand{\mguts}{{\ifmmode M_{GUT}^2  \else $M_{GUT}^2$\fi}}
\newcommand{\mze} {{\ifmmode m_0        \else $m_0$\fi}}
\newcommand{\mha}{{\ifmmode m_{1/2}    \else $m_{1/2}$\fi}}
\newcommand{\mb} {{\ifmmode m_{b}    \else $m_{b}$\fi}}
\newcommand{\mt} {{\ifmmode m_{t}    \else $m_{t}$\fi}}
\newcommand{\mts} {{\ifmmode m_{t}^2    \else $m_{t}^2$\fi}}
\newcommand{\tb} {{\ifmmode \tan\beta  \else $\tan\beta$\fi}}
\newcommand{\msoten} {{\ifmmode M_\rST         \else $M_\rST$          \fi}}
\newcommand{\rPL}  {{\rm Planck}}
\newcommand{\mplanck} {{\ifmmode M_\rPL         \else $M_\rPL$          \fi}}
\newcommand{\rST}  {{\rm SO(10)}}
\newcommand{\ai}   {{\ifmmode \alpha_i      \else $\alpha_i$       \fi}}
\newcommand{\aii}  {{\ifmmode \alpha_i^{-1} \else $\alpha_i^{-1}$  \fi}}
\newcommand{\DRbar}{{\ifmmode \overline{DR} \else $ \overline{DR}$ \fi}}
\newcommand{\tal}  {{\ifmmode \tilde{\alpha} \else $\tilde{\alpha}$ \fi}}
\newcommand {\tabs}[1]{\multicolumn{1}{c}{\mbox{\hspace{#1}}}}

\hyphenation{multi-pli-ci-ties}
\hyphenation{Su-per-sym-me-try}
\hyphenation{cor-rections}
\hyphenation{pa-ra-me-ter}
\hyphenation{pa-ra-me-ters}
\newcommand{\mtau}{{\ifmmode m_{\tau}  \else $m_{\tau}$\fi}}
\newcommand{\dpp}{{\ifmmode \delta_{pert} \else $\delta_{pert}$\fi}}
\newcommand{\dnp}{{\ifmmode\delta_{non-pert}\else$\delta_{non-pert}$\fi}}
\newcommand{\dew}{{\ifmmode \delta_{\rm EW}\else $\delta_{\rm EW}$\fi}}
\newcommand{\rt}{{\ifmmode R_{\tau}  \else
                 $R_{\tau} $\fi}}
\newcommand{\rz}{{\ifmmode R_{Z}  \else
                 $R_{Z} $\fi}}
\newcommand{\into}{\rightarrow}
\newcommand{\SM}{Standard Model}
\newcommand{\swb}{{\ifmmode \sin^2\theta_{\overline{MS}}
                     \else $\sin^2\theta_{\overline{MS}}$\fi}}
\newcommand{\cwb}{{\ifmmode \cos^2\theta_{\overline{MS}}
                     \else $\cos^2\theta_{\overline{MS}}$\fi}}
\def\ai{\alpha_i}
\def\aii{\alpha_i^{-1}}
\def\rZ{{\rm Z}}
\def\rW{{\rm W}}
\def\rG{{\rm GUT}}
\def\rt{{\rm threshold}}
\def\rS{{\rm SUSY}}
\def\rH{{\rm Higgs}}
\def\rF{{\rm Fam}}
\def\MG{M_\rG}
\def\MS{M_\rS}
\def\MZ{M_\rZ}
\def\MW{M_\rW}
\def\Mt{M_\rt}
\def\MSbar{{\overline{MS}}}
\def\DRbar{{\overline{DR}}}                                            \newcommand{\Z}{\mbox{$Z^{0}$}}
\newcommand{\WW}{\mbox{$W^{\pm}$}}
\newcommand{\EE}{\mbox{$e^{+}e^{-}$}}
\newcommand{\MM}{\mbox{$\mu^{+}\mu^{-}$}}
\newcommand{\TT}{\mbox{$\tau^{+}\tau^{-}$}}
\newcommand{\GSW}{\mbox{\sc GSW}}
\newcommand{\QCD}{\mbox{\sc QCD}}
\newcommand{\QED}{\mbox{\sc QED}}
\newcommand{\SLC}{\mbox{\sc SLC}}
\newcommand{\LEP}{\mbox{\sc LEP}}
\newcommand{\CERN}{\mbox{\sc CERN}}
\newcommand{\PETRA}{\mbox{\sc PETRA}}
\newcommand{\DESY}{\mbox{\sc DESY}}
\newcommand{\SLAC}{\mbox{\sc SLAC}}
\newcommand{\ALEPH}{\mbox{\sc ALEPH}}
\newcommand{\DELPHI}{\mbox{\sc DELPHI}}
\newcommand{\OPAL}{\mbox{\sc OPAL}}
\newcommand{\DEG}{\mbox{$^{\circ}$}}
\newcommand{\DELSIM}{\mbox{\tt DELSIM}}
\newcommand{\DELANA}{\mbox{\tt DELANA}}
\newcommand{\SUSY}{\mbox{\sc SUSY}}
\newcommand{\GUT}{\mbox{\sc GUT}}
\newcommand{\LL}{{\ifmmode {\cal L} \else ${\cal L}$\fi}}
\newcommand{\hz}{{\ifmmode {\rm Hz} \else ${\rm Hz}$\fi}}
\newcommand{\khz}{{\ifmmode {\rm kHz} \else ${\rm kHz}$\fi}}
\newcommand{\mhz}{{\ifmmode {\rm mHz} \else ${\rm mHz}$\fi}}
\newcommand{\as}{{\ifmmode \alpha_s  \else $\alpha_s$\fi}}
\newcommand{\asmz}{{\ifmmode \alpha_s(M_Z) \else $\alpha_s(M_Z)$\fi}}
\newcommand{\astau}{{\ifmmode \alpha_s(M_{\tau})
                       \else $\alpha_s(M_{\tau})$\fi}}
\newcommand{\ca}{{\ifmmode C_a  \else $C_a$\fi}}
\newcommand{\tf}{{\ifmmode T_{\mbox{\scriptsize Fermion}}
           \else $T_{\mbox{\scriptsize Fermion}}$\fi}}
\newcommand{\ts}{{\ifmmode T_{\mbox{\scriptsize Scalar}}
           \else $T_{\mbox{\scriptsize Scalar}}$ \fi}}
\newcommand{\mhiggs}{{\ifmmode M_{\mbox{\scriptsize Higgs}}
           \else $M_{\mbox{\scriptsize Higgs}}$\fi}}
\newcommand{\mthres}{{\ifmmode M_{\mbox{\scriptsize threshold}}
           \else $M_{\mbox{\scriptsize threshold}}$ \fi}}
\newcommand{\msbar}{{\ifmmode \overline{MS} \else $\overline{MS}$\fi}}
\newcommand{\drbar}{{\ifmmode \overline{DR} \else $\overline{DR}$\fi}}
\newcommand{\lamms}{{\ifmmode \Lambda_{\overline{MS}}
                       \else $\Lambda_{\overline{MS}}$\fi}}
\newcommand{\PL}{Phys. Lett.}
\newcommand{\PRL}{Phys. Rev. Lett.}
\newcommand{\NP}{Nucl. Phys.}
\newcommand{\rr}{{{\ifmmode {\cal R}_2 }\else ${\cal R}_2 $\fi}}
\newcommand{\rrr}{{{\ifmmode {\cal R}_3 }\else ${\cal R}_3 $\fi}}
\newcommand{\rrrr}{{{\ifmmode {\cal R}_4 }\else ${\cal R}_4 $\fi}}
\newcommand{\jdd}{{{\ifmmode {\cal D}_2 }\else ${\cal D}_2 $\fi}}
\newcommand{\jddd}{{{\ifmmode {\cal D}_3 }\else ${\cal D}_3 $\fi}}
\newcommand{\jdddd}{{{\ifmmode {\cal D}_4 }\else ${\cal D}_4 $\fi}}
\newcommand{\rrre}{{{\ifmmode {\cal R}_3^{E0}}\else ${\cal R}_3^{E0}$\fi}}
\newcommand{\rrrp}{{{\ifmmode {\cal R}_3^P}\else ${\cal R}_3^P$\fi}}
\newcommand{\jdde}{{{\ifmmode {\cal D}_2^{E0}}\else ${\cal D}_2^{E0} $\fi}}
\newcommand{\jddp}{{{\ifmmode {\cal D}_2^P}\else ${\cal D}_2^P$\fi}}
\newcommand{\ycut}{{{\ifmmode y_{cut} }\else $y_{cut}$\fi}}
\newcommand{\ymin}{{{\ifmmode y_{min} }\else $y_{min}$\fi}}
\newcommand{\sph}{{{\ifmmode {\cal S} }\else ${\cal S} $\fi}}
\newcommand{\apl}{{{\ifmmode {\cal A} }\else ${\cal A} $\fi}}
\newcommand{\thr}{{{\ifmmode {\cal T} }\else ${\cal T} $\fi}}
\newcommand{\obl}{{{\ifmmode {\cal O} }\else ${\cal O} $\fi}}
\newcommand{\cpa}{{{\ifmmode {\cal C} }\else ${\cal C} $\fi}}
\newcommand{\eec}{{{\ifmmode {\cal E}{\cal E}{\cal C} }\else
${\cal E}{\cal E}{\cal C}$\fi}}
\newcommand{\aeec}{{{\ifmmode {\cal A}{\cal E}{\cal E}{\cal C} }
\else ${\cal A}{\cal E}{\cal E}{\cal C}$\fi}}
\newcommand{\hjm}{{\ifmmode {\bf M^2_{high}}
                   \else   ${\bf M^2_{high}}$\fi}}
\newcommand{\ljm}{{\ifmmode {\bf M^2_{low}}
                   \else   ${\bf M^2_{low}}$\fi}}
\newcommand{\djm}{{\ifmmode {\bf M^2_{diff}}
                   \else   ${\bf M^2_{diff}}$\fi}}
\newcommand{\hjmt}{{\ifmmode {\bf M({\cal T})^2_{high}}
                    \else   ${\bf M({\cal T})^2_{high}}$\fi}}
\newcommand{\ljmt}{{\ifmmode {\bf M({\cal T})^2_{low}}
                    \else   ${\bf M({\cal T})^2_{low}}$\fi}}
\newcommand{\djmt}{{\ifmmode {\bf M({\cal T})^2_{diff}}
                    \else   ${\bf M({\cal T})^2_{diff}}$\fi}}
\newcommand{\djr}{{{\ifmmode {\bf {\cal D}_2}\else ${\bf {cal D}_2}$\fi}}}
\newcommand{\ma}{{{\ifmmode {\bf {\cal M}_{Major}}
\else ${\bf {\cal M}_{Major}}$\fi}}}
\newcommand{\mi}{{{\ifmmode {\bf {\cal M}_{Minor}}
\else ${\bf {\cal M}_{Minor}}$\fi}}}
\newcommand{\ps}{{\mbox{\bf PS}}}
\newcommand{\me}{\mbox{\bf ME}}
\newcommand{\ha}{{{\ifmmode {\frac{1}{2}}\else ${\frac{1}{2}}$\fi}}}
\newcommand{\Q}{\widetilde m_Q^2}
\newcommand{\U}{\widetilde m_U^2}
\newcommand{\D}{\widetilde m_D^2}
\newcommand{\E}{\widetilde m_E^2}
\renewcommand{\L}{\widetilde m_L^2}
\newcommand{\q}[2]{\widetilde m_{Q_{#1#2}}^2}
\renewcommand{\u}[2]{\widetilde m_{U_{#1#2}}^2}
\renewcommand{\d}[2]{\widetilde m_{D_{#1#2}}^2}
\renewcommand{\l}[2]{\widetilde m_{L_{#1#2}}^2}
\newcommand{\e}[2]{\widetilde m_{E_{#1#2}}^2}
\newcommand{\k}[2]{K_{#1#2}{}}
\renewcommand{\a}[1]{\widetilde\alpha_{#1}}
\newcommand{\hu}{\widetilde h_U{}}
\newcommand{\hd}{\widetilde h_D{}}
\newcommand{\hdd}{\widetilde h_{D_{23}}}
\newcommand{\he}{\widetilde h_E{}}
\newcommand{\au}[2]{A_{U_{#1#2}}}
\newcommand{\ad}[2]{A_{D_{#1#2}}}
\renewcommand{\ae}[2]{A_{E_{#1#2}}}
\newcommand{\htop}{\widetilde h_t}
\newcommand{\hbot}{\widetilde h_b}
\newcommand{\htau}{\widetilde h_{\tau}}
\renewcommand{\atop}[2]{A_{t_{#1#2}}}
\newcommand{\abot}[2]{A_{b_{#1#2}}}
\newcommand{\atau}[2]{A_{\tau_{#1#2}}}
\renewcommand{\arraystretch}{1.7}
\newcommand{\be}{\small\begin{eqnarray*}}
\newcommand{\ee}{\end{eqnarray*}}
\renewcommand{\d}[2]{\widetilde m_{D_{#1#2}}^2}
\renewcommand{\u}[2]{\widetilde m_{U_{#1#2}}^2}
%


\pagestyle{empty}

\begin{flushright}
\vspace{-2.3cm}
        IEKP-KA/97-03    \\
hep-ph/9705309   \\
        March, 1997       \\
\end{flushright}

\vspace{1.cm}

\begin{center}
{\bf \LARGE     In Search of SUSY \\}

\vspace{1.0cm}
{\bf    W.  de Boer\footnote{E-mail: wim.de.boer@cern.ch\\
Invited talk at the EPIPHANY Workshop, Cracow, Jan. 1997.}\\}
\baselineskip=13pt
{\it Inst.\ f\"ur Experimentelle Kernphysik\\ Universit\"at  Karlsruhe   \\
  Postfach 6980\\
  D-76128 Karlsruhe}
\baselineskip=12pt
\vspace{2.0cm}
{\bf    ABSTRACT}
\end{center}
\vspace{0.3cm}

\begin{center}\parbox{13cm}{
Electroweak precision tests of the SM and MSSM as well as
 Searches for Supersymmetric Particles and Higgs
bosons at LEP II and their significance within the MSSM
are discussed.}
\end{center}

\pagestyle{plain}
\pagenumbering{arabic}
\setcounter{page}{1}
\tableofcontents
\clearpage
%
\section{Introduction}

Although at present the Standard Model (SM) shows good agreement with all 
available data, many questions can only be answered by assuming 
new physics beyond the SM.
An excellent candidate for new physics is the
the supersymmetric extension of the SM (MSSM), which 
was found to  describe the electroweak data equally well.
In  addition the MSSM allows 
\begin{itemize}
\item Unification of the gauge coupling constants;
\item Unification of the Yukawa couplings;
\item Natural occurrence of the Higgs mechanism  at a low scale;
\item Cancellation of the quadratic divergences in the 
radiative corrections of the SM
\item Relic abundance of dark matter.
\end{itemize}
After the discovery that unification within the SM is
 excluded by the precise measurements of the 
coupling constants at LEP I\cite{ekn,abf,lanluo},
a flood of papers on these subjects have 
emerged. Some recent 
contributions of the groups involved are 
given in refs. 
\cite{MSSM}\nocite{rrb,car,acpz,nan,bek,ir,arn}-\cite{roskane}
It is surprising that one can find  a region of parameter space
 within the
{\it minimal}~ SUSY model, where all the
independent constraints mentioned above can be fulfilled 
simultaneously.

The paper has been organized as follows: 
first the electroweak precision tests of the SM and MSSM
are discussed, followed by   the corresponding 
restrictions on the MSSM parameter space,
both from the searches and the unification conditions.

\section{Electroweak Precision Tests of the SM and MSSM}

In this section  an equivalent analysis of all electroweak data,
both in the SM and its supersymmetric extension, is described
using all actual electroweak data from Tevatron, 
LEP and SLC \cite{EWWG},
the measurement of
$\frac{BR(b\rightarrow s\gamma)}{BR(b\rightarrow ce\bar\nu)}$ 
from CLEO \cite{cleo}
and limits on the masses of supersymmetric
particles.
The observed $b\rightarrow s\gamma$ decay rate 
is $30\%$  below the SM prediction, while the decay 
 $Z_0\rightarrow b\overline{b}$ is about 1.8$\sigma$ above the SM prediction.
In the MSSM light stops and light  charginos increase 
$R_b$\cite{yel1}
\nocite{boufi,chan2,ell1,garc3,kan1,kan2,garcia}-\cite{garcia2} 
 and decrease the  
$b\rightarrow s\gamma$ rate, so both observations can be brought into 
agreement with the MSSM for the same region of parameter space.
However, as will be shown,  the resulting    
$\chi^2$ value for the MSSM fits is 
only marginally lower. In addition,  the splitting in the stop 
sector has to be  unnaturally high, 
so it remains to be seen if these effects
are real or due to a fluctuation.
Further details of the  procedure and extensive references are given
  elsewhere\cite{hollik1}.
%
%
\subsection{\it Standard Model Fits}
The SM cross sections and asymmetries 
are completely determined by $M_Z,m_t,m_H,G_F,\alpha,\alpha_s$.
From the combined CDF and D0 data $m_t$ has 
 been determined to be $175\pm6$ GeV\cite{top},
so the parameters with the largest uncertainties are $m_H$ and $\alpha_s$.
The error on the finestructure constant $\alpha$ is limited by the uncertainty in the
hadronic cross section in $e^+e^-$ annihilation at low energies, 
which is used to determine the vacuum polarization contributions to $\alpha$.
The error was taken into account by considering $\alpha$ to be a free parameter in the fit
and constraining it to the value $1/\alpha=128.89\pm0.09$\cite{jegerlehner}.
If this error is not taken into account, the error on the Higgs mass 
is underestimated by 30\%. 
Using the input values discussed in the introduction yields:
\begin{eqnarray*}
\alpha_s&=&0.120\pm0.003\\
m_t&=&172.0^{+5.8}_{-5.7}~{\rm GeV}\\
m_H&=&141^{+140}_{-77}~{\rm GeV} 
\end{eqnarray*}
Minor deviations from the EWWG fit results\cite{EWWG}
 are due to the incorporation
of the $b\rightarrow s\gamma$ data from CLEO\cite{cleo},
 which are  important for
the MSSM fits described below.
From the SM fit parameters one can derive the value of the
electroweak mixing parameter in the $\overline{MS}$ scheme:
$sin^2\theta_{\overline{MS}}=0.2316\pm0.0004,$
which is within errors  equal to 
$\sin^2\Theta_{eff}^{lept}$.
The main contributions to the
$\chi^2/d.o.f=18.5/15$ originate from $\sin^2\Theta_{eff}^{lept}$ 
from SLD  
($\Delta\chi^2=4.9$), $R_b$ ($\Delta\chi^2=3.1$) 
and $A_{FB}^b$  ($\Delta\chi^2=3.5$), but  the overall SM 
agreement is good: the $\chi^2/d.o.f.$=18.5/15  corresponds 
to a probability of 24\%.

\begin{figure}[t]
 \begin{center}
\parbox{0.65\textwidth}{
  \leavevmode
  \epsfxsize=0.65\textwidth
  \epsffile{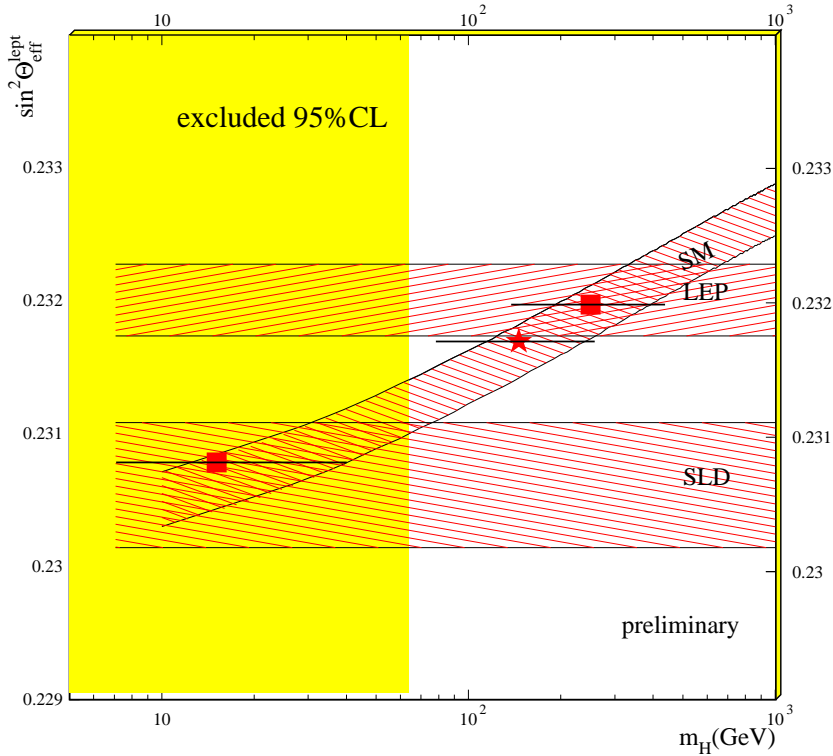}
}
\hfill
\parbox{0.3\textwidth}{
\caption{\label{sintw}
Dependence of the SM $\sin^2\Theta_{eff}^{lept}$ on the Higgs mass. 
The top mass $m_t=175\pm 6$~GeV was varied within its error, 
as shown by the dashed band labelled SM. 
The SLD and the LEP measurements of  $\sin^2\Theta_{eff}^{lept}$ 
are also shown as horizontal bands.
The SLD value yields a Higgs mass below the recents limits 
 by direct Higgs searches at LEP
(shaded area).
}}
\end{center}\vspace{-0.95cm}
\end{figure}

\begin{figure}[t]
 \begin{center}
\parbox{0.67\textwidth}{
  \leavevmode
  \epsfig{file=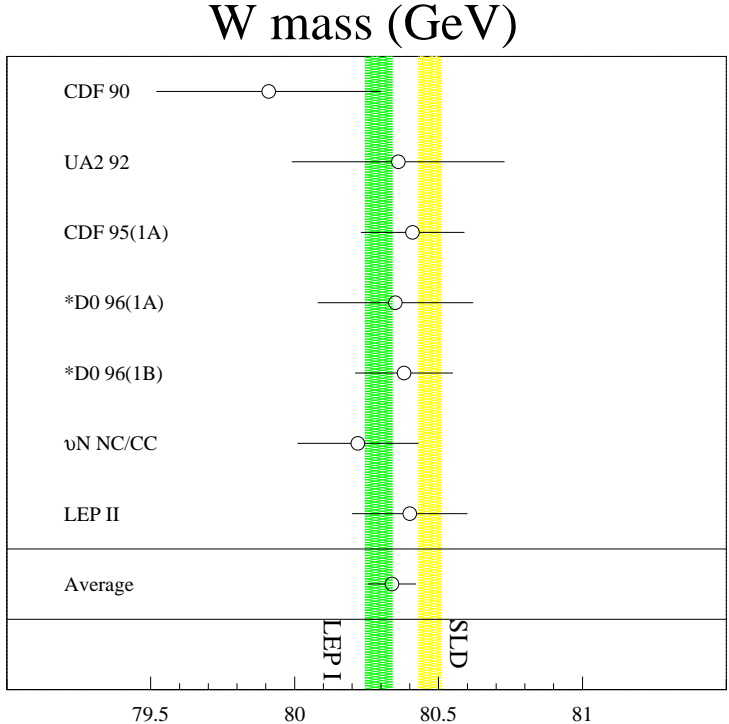,width=0.67\textwidth}
}
\hfill
\parbox{0.3\textwidth}{
\caption[]{\label{mw}
A compilation of the W-masses. The vertical lines indicate the predictions
from the LEP I and   SLD electroweak data determining 
$\sin^2\Theta_{eff}^{lept}$ and the data points represent 
the various direct measurements of $M_W$. 
} }
\end{center}\vspace{-1cm}
\end{figure}

The low value of $\sin^2\Theta_{eff}^{lept}$ from SLD as compared
to the LEP value yields a Higgs mass  below the lower limit on
   the SM Higgs mass from direct searches, as demonstrated
in   fig. \ref{sintw}.
The LEP data alone without SLD yield $m_H\approx240$~GeV, 
while $\sin^2\Theta_{eff}^{lept}$
from SLD corresponds to $m_H\approx15$~GeV,
as indicated by the squares in fig. \ref{sintw}.
The latter value is excluded by the 95\% C.L. 
 lower limit of 63.9~GeV from 
the  LEP experiments~\cite{rev96,higgslim}.
The different values of $\sin^2\Theta_{eff}^{lept}$ 
from LEP and SLD translate into different
predictions for $M_W$, as shown in fig. \ref{mw}. 
The present $M_W$ measurements,
including the preliminary value from the LEP II measurements\cite{EWWG1}
lie in between these  predictions.

\subsection{\it MSSM Fits and Comparison with the SM}

\begin{figure}[t]
 \begin{center}
\parbox{0.67\textwidth}{
  \leavevmode
  \epsfig{file=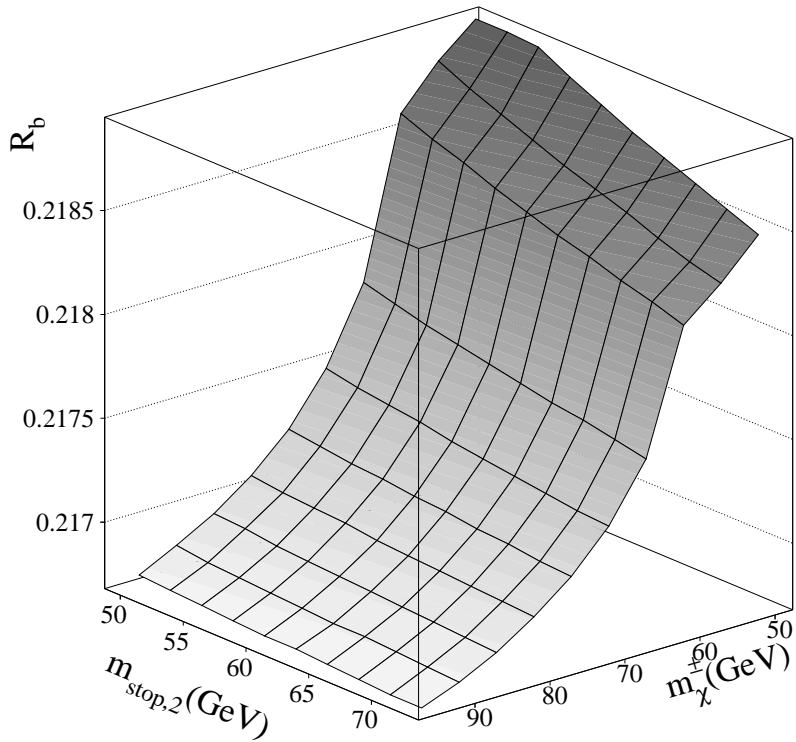,width=0.67\textwidth}
}
\hfill
\parbox{0.3\textwidth}{
\caption[]{\label{rb}
$R_b$ as function of the stop and chargino masses for $\tb=1.6$.
 The experimental value $R_b=0.2178\pm 0.0011$ is clearly 
above the SM value of 0.2158 and can be obtained for 
light charginos and stops.
}}
\end{center}\vspace{-1cm}
\end{figure}

As mentioned in the introduction, 
the MSSM can increase the value of $R_b$,
which experimentally is slightly above the SM value.
The major additional contributions originate 
from vertex contributions with light 
charginos and light right handed stops in the low $\tan\beta$ scenario
and light higgses for large $\tan\beta $ values. 
Since the large $\tan\beta$
scenario does not improve $R_b$ significantly\cite{hollik1}, 
it will not be discussed here anymore.
 The $R_b$ dependence on  chargino and stop masses is shown in
fig. \ref{rb}.
 The experimental value $R_b=0.2178\pm 0.0011$ is clearly 
above the SM value of 0.2158 and can be obtained for   
charginos around 85 GeV  and the lightest stop mass around 50 GeV 
(best fit results,\cite{hollik1}), although the second stop
mass has to be heavy, i.e. well above $m_t$. 

As will be discussed in the next section, such a large splitting in the
 stop sector is difficult to obtain in the MSSM, if one requires 
unification of the left and right-handed stop squarks at the GUT scale. 
Final analysis of available LEP data  will teach of the 
present preliminary value of $R_b$ will indeed stay above the SM value.

%
%
%
\begin{figure}[t]
 \begin{center}
\parbox{0.63\textwidth}{
  \leavevmode
  \epsfig{file=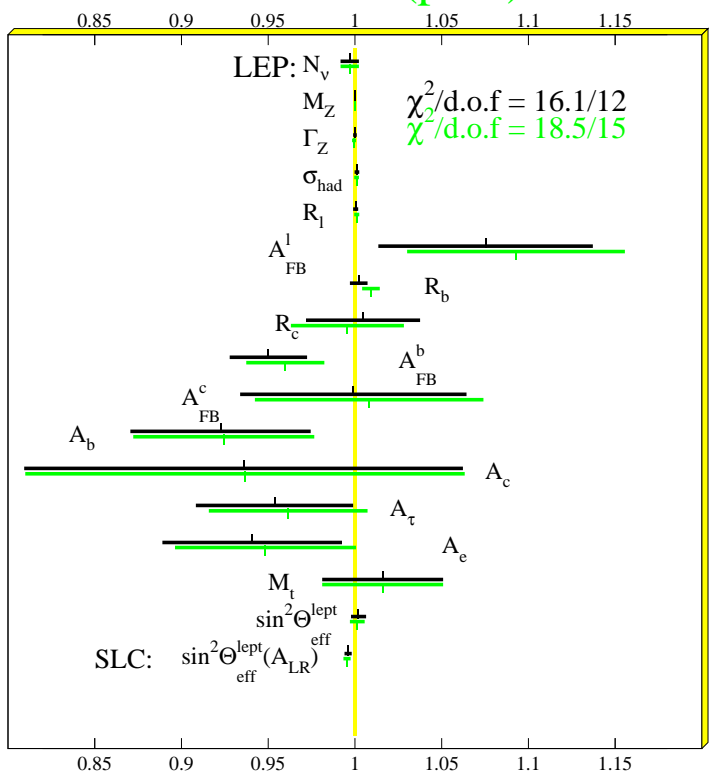,width=0.67\textwidth}
}
\hfill
\parbox{0.34\textwidth}{
\caption{\label{\figV}
Fit results normalized to the SM- and MSSM ($\tan\beta=1.6$) values.
The difference in $\chi^2/d.o.f$ between the SM and MSSM originates 
mainly from $\sin^2\Theta_{eff}^{lept},$ $A_{FB}^b$ and $R_b$.
}}
\end{center}\vspace{-1.5cm}
\end{figure}

The fit results  are 
compared with the Standard Model fits   in
fig.~\ref{\figV}. The  Standard Model
$\chi^2/d.o.f.=18.5/15$ corresponds to a pro\-bability of 24\%,
the MSSM  $\chi^2/d.o.f.=16.1/12$  to a probability of 19\%. 
In counting the d.o.f the insensitive (and fixed) parameters were 
ignored \cite{hollik1}.

\begin{figure}[t]
 \begin{center}
\parbox{0.63\textwidth}{
  \leavevmode
  \epsfig{file=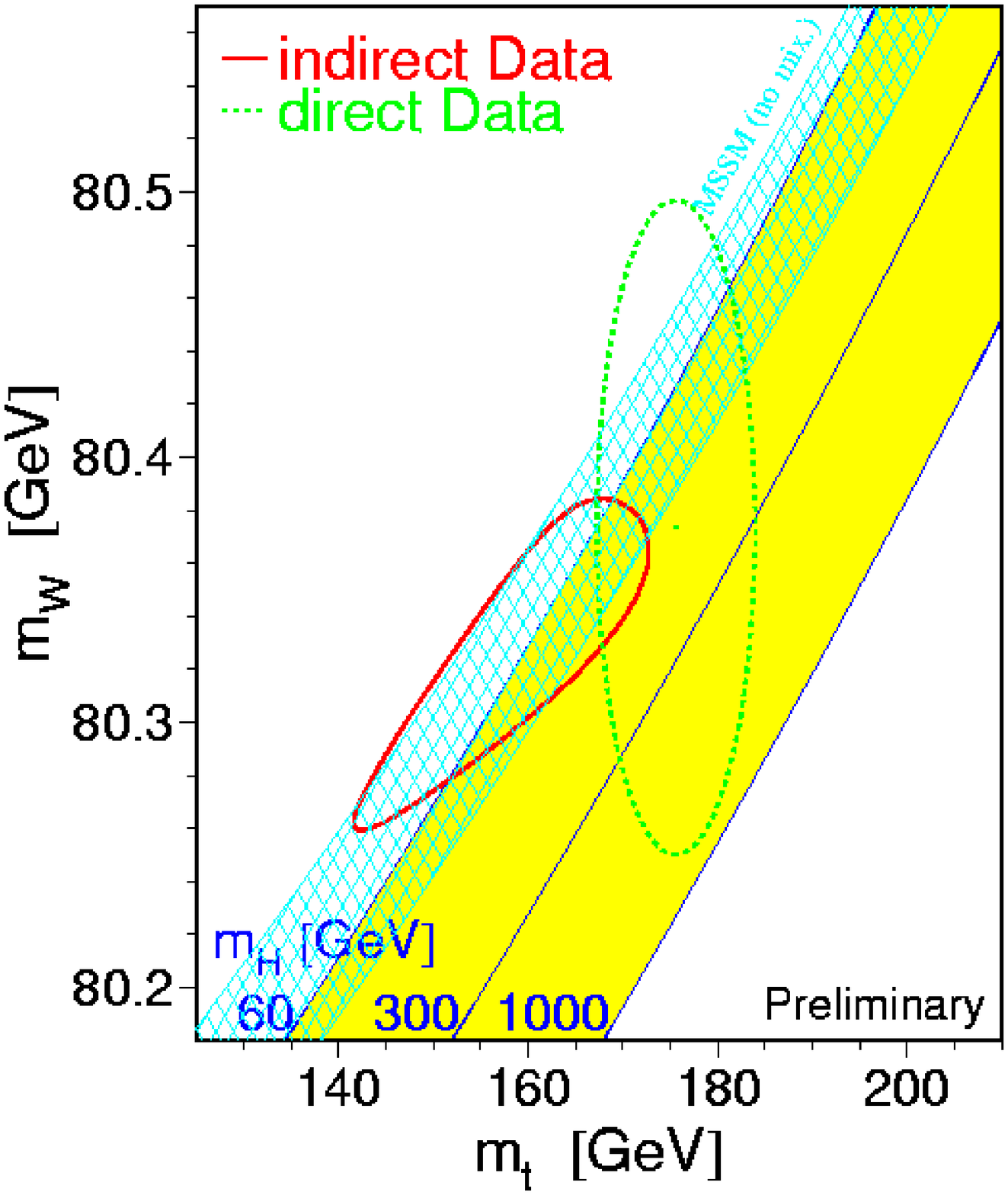,width=0.67\textwidth}
}
\hfill
\parbox{0.33\textwidth}{
\caption[]{\label{mwmt}
$m_W$  and $m_t$ from direct (Tevatron and LEP II)  and indirect 
measurements  in comparison with the SM (shaded area) and MSSM (crossed area)  predictions.
The uncertainty from the SM prediction  originates from the unknown Higgs mass,
while for the MSSM it is mainly the uncertainty from the stop mass, since
the Higgs mass is quite well predicted in the MSSM. The highest $m_W$ mass
is obtained for the lightest stop mass.
}}
\end{center}\vspace{-1cm}
\end{figure}

It is interesting to note that the predicted value of $m_W$ 
tends to be  higher in the MSSM than in the SM, especially for light stops,
as shown in fig. \ref{mwmt}.
 
Another interesting point are the    $\alpha_s(M_Z)$ values.
An increase in $R_b$ implies in increase in the total width of the $Z^0$ boson, which
can be compensated by a decrease in the QCD corrections, i.e. $\alpha_s$.
However, since $R_b$ is only marginally above the SM value,
the fitted value of $\alpha_s(M_Z)$ between SM and MSSM is within
the error bars. 
Note that the $\alpha_s$ crisis has disappeared after the LEP value from the total
cross section came down and the value from both lattice calculations and deep 
inelastic scattering went up\cite{schmelling}.


%
%


\section{The Minimal SuperSymmetric Model (MSSM)}\label{mssm}

Supersymmetry presupposes a symmetry between fermions and bosons,
which can only be realized in nature by assuming for every particle of the SM
with spin $j$ a supersymmetric partner (sparticles) with spin $j-1/2$. These
spartners must have the same mass and couplings as the particles,
if supersymmetry is an exact symmetry in nature.  
However, since the sparticles have not been observed sofar, supersymmetry 
must be broken. The MSSM can be obtained from the SM by replacing the
known fields with the superfields, which include the spin 0 sfermions
and the spin 1/2 gauginos. 

In addition supersymmetry requires 
 two complex SU(2) doublets for the Higgs sector 
instead of only one in the SM.
The reasons are twofold: a) in the SM one can give mass to the
down-type quarks and leptons by using the complex conjugate of the Higgs 
doublet. Since the Higgses in the MSSM are part of the bosonic fields, one
cannot just take the complex conjugate of just a part of the superfield
structure, so one needs separate Higgs doublets for up- and down-type
fermions.
b) the superpartners of the Higgses are fermions, which contribute the
triangle anomalies, unless the total hypercharge equals zero.
This requires the introduction of two SU(2) Higgs doublets with
opposite hypercharge.

Since the top quark is much heavier than the bottom quark, the 
Yukawa corrections for the two mass terms of the two   
Higgs doublets are very different,
thus breaking the symmetry between them. These radiative
 corrections  automatically lead
to the Higgs mechanism of spontaneous electroweak symmetry breaking
at a scale far below the unification scale, as discussed in many 
reviews\cite{MSSM}.

Another difference between  the interactions in the SM and MSSM
arises from the triple vertices: in the SM a spin 1/2 fermion cannot
couple to two other fermions, since this would violate conservation
of angular momentum (or more general Lorentz invariance).
For spin 0 particles such triple vertices are allowed,
so  fermions can couple to a sfermion and a fermion!
%
Such vertices  with three fermions violate
lepton and/or baryon number.
They can be avoided in the MSSM by introducing an 
additional multiplicative quantum number, 
called $R-parity$, defined as:
\begin{equation}
R=(-1)^{3(B-L)+2S}. \label{par}
\end{equation}
This quantity is +1 for SM particles and -1 for the supersymmetric
partners, because of the change in the spin S.
$R$-parity consercation  forbids the coupling of a fermion to a 
sfermion and fermion,
since the final state would have $R=-1~{.}~ {+1}=-1$, thus
eliminating the dangerous baryon - and lepton number violating
vertices.
It is usually assumed
that  $R$-parity is conserved
exactly, since
the experimental limits on the $R$-parity violating couplings are
very severe.
R-parity conservation implies that:
\begin{itemize}
\item
sparticles can only be produced   in pairs
\item
the lightest supersymmetric particle
is stable, since its decay into normal matter would violate R-parity.
\item
the interactions of particles and sparticles
can be different.  For example, the photon couples to
electron-positron pairs, but the photino does
not couple to
selectron-spositron pairs, since in the latter case
the R-parity would change from -1 to +1.
In other words, each triple vertex must have   2 sparticles attached to
it, thus forbidding the triple vertices in which a fermion couples to
a sfermion and another fermions.
\end{itemize}


Obviously SUSY cannot be an exact symmetry of nature; 
else the supersymmetric partners would have the
same mass as the normal particles.
In the absence of a fundamental understanding of the origin of 
supersymmetry breaking one considers all breaking terms, 
which do not  introduce quadratic divergences. 
This cancellation between fermions and bosons in the 
loop corrections is one of the great advantages of the MSSM,
since it allows one to calculate radiative corrections 
up to the unification scale without divergences.

The breaking terms consist of the gaugino mass terms, the
scalar mass terms, the trilinear (A-term) interactions 
amongst the scalars and the analogous bilinear (B-term) 
interactions\cite{soft}.

If one assumes that SUSY  is broken due to the {\it universal}
 gravitational interactions
one needs only a {\it few} independent SUSY breaking parameters
at the unification scale:
 a common mass
$m_{1/2}$ for the gauginos,
 a common mass $m_0$
for the scalars,
 a common trilinear interaction $A_0$ and a bilinear coupling $B_0$.

In addition to these soft breaking terms one needs to specify the 
ratio $\tan\beta$ of the two Higgs VEVs and a supersymmetric 
Higgsino mixing parameter $\mu$. 
The minimization conditions of the Higgs potential
requiring a non-trivial minimum for electroweak 
symmetry breaking\cite{ewbr} yields a
relation between the
bilinear coupling $B_0$ and  $\tan\beta$ and 
determines the value of $\mu^2$,
so finally the SUSY mass spectrum in this 
supergravity inspired scenario
is determined by the following parameters:
\bqa
m_0,~ m_{1/2},~\tan\beta,~A_0,~sign(\mu)\label{free}
\eqa
As will be shown in the next section, $\tb$ has only two solutions
from   the known top mass, while  $sign(\mu)$ and $A_0$ do 
not influence the mass
spectrum strongly (except for the mixing in the stop sector, 
which can change the lightest Higgs mass by  10-15 GeV), 
so the main variables for the prediction of the SUSY mass
spectrum are $m_0$ and $m_{1/2}$.
The various MSSM masses and couplings have to be evolved via the 
 renormalization group equations (RGE) from their common 
value at the unification scale to the electroweak scale. This involves
 solving typically 26 coupled differential equations with common values
 as boundary conditions  at $\mgut$ $(t=\ln ({M/\mgut})^2=0)$:
\bqa
 {\rm scalars:}&&
 \tilde{m}^2_Q=\tilde{m}^2_U=
\tilde{m}^2_D=\tilde{m}^2_L=\tilde{m}^2_E=
m_0^2;\\
 {\rm gauginos:}&&
 M_i=m_{1/2}, \ \ \ i=1,2,3;\\
 {\rm couplings:}&&
 \tilde{\alpha}_i(0)=\tilde{\alpha}_{GUT},\ \ \ i=1,2,3 .\eqa
 Here $M_1$, $M_2$, and $ M_3$ are the  gauginos masses of the 
$U(1)$, $SU(2)$ and $SU(3)$ groups. 
One has, however, to take into account the mixing
between various states.

\subsection{Gaugino-Higgsino Mass Terms: Charginos and Neutralinos}
Gauginos and Higgsinos both have spin $j=1/2$, 
so the mass eigenstates can
be different from the interaction eigenstates 
because of the non-diagonal mass
terms. The partners of the two neutral gauge bosons 
and two neutral Higgs bosons are
the four neutralinos $\tilde{\chi}^0_i~( i=1,4)$ 
after mixing; correspondingly, the
 charginos $\tilde{\chi}^\pm_i~(i=1,2)$  are mixtures of 
the wino  and charged higgsino.
The neutralino mixing  is described by the following mass matrix: 
{\small
\begin{equation}
M^{(0)}=\left(
\begin{array}{cccc}
M_1 & 0 & -M_Z\cos\beta \sin_W & M_Z\sin\beta \sin_W \\
0 & M_2 & M_Z\cos\beta \cos_W   & -M_Z\sin\beta \cos_W  \\
-M_Z\cos\beta \sin_W & M_Z\cos\beta \cos_W  & 0 & -\mu \\
M_Z\sin\beta \sin_W & -M_Z\sin\beta \cos_W  & -\mu & 0
\end{array} \right).\label{neumat}
\end{equation}
}
The physical neutralino masses  $M_{\tilde{\chi}_i^0}$
are obtained as eigenvalues of this matrix after diagonalization.
For charginos one has similarly:
{\small
\begin{equation}
M^{(c)}=\left(
\begin{array}{cc}
M_2 & \sqrt{2}M_W\sin\beta \\ \sqrt{2}M_W\cos\beta & \mu
\end{array} \right).\label{chamat}
\end{equation}
}
The $M_1$ and $M_2$ terms are the gaugino masses at low energies.
They are linked to their coomon values at the GUT scale ($m_{1/2})$
by  the RGE group equations.
Numerically one finds at the weak scale: 
\bqa
M_3(\tilde{g})&\approx& 2.7m_{1/2},\\ 
 M_2(M_Z)&\approx& 0.8m_{1/2},\\  
 M_1(M_Z)&\approx& 0.4m_{1/2},\\
  \mu(M_Z)&\approx&0.63 \mu(0).
\label{gaugino}
\eqa
Since the gluinos obtain corrections from
the strong coupling constant $\alpha_3$, they grow
heavier than the gauginos of the $SU(2)\otimes U(1)$ group.

In the case favoured by the fit discussed
below one finds $\mu >>M_2> M_W$, in which case the
charginos eigenstates are approximately
$M_2$ and $\mu $ and the four neutralino mass eigenstates are 
$|M_1|,|M_2|, |\mu|,$ and $|\mu|$,
respectively.
In other words, the  neutralinos and charginos do not  mix strongly,
so the lightest chargino is wino-like, while the 
 the LSP is bino-like,
which has consequences for dark matter searches.

\subsection{Squark and Slepton Masses}

The non-negligible Yukawa couplings cause a mixing between the electroweak
eigenstates and the mass eigenstates of the third generation particles.  The
mixing matrix for the stopsector is:
{\small
\begin{equation} \label{stopmat}
\left(\begin{array}{cc} \tilde m_{tL}^2& m_t(A_t-\mu\cot \beta ) \\
m_t(A_t-\mu\cot \beta ) & \tilde m_{tR}^2 \end{array}  \right).
\nonumber
\end{equation}
}
The  mass eigenstates are  the eigenvalues of this  matrix.
Similar matrices exist for   sbottom and stau, except
that in the off-diagonal elements $m_t$ is replaced by $m_{b(\tau)}$
and $\cot\beta$ is replaced
by $\tan\beta$, so the mixing effects are smaller, 
unless $\tan\beta$ is large. For the first and second generation the 
mixing can be neglected, since the off-diagonal terms
are proportional to the quark masses of the first and second generation.

 The squark and slepton masses are assumed to all have 
  the same value at the GUT scale. However, in contrast to the
sleptons, the squarks get 
radiative corrections from virtual gluons
which make them heavier than the
sleptons at low energies.

\subsection{CMSSM and $R_b$}

An increase in $R_b$ requires
one (mainly right handed) stop to be light and 
the other one to be heavy (see previous section).
If both would be light, then all other squarks are likely to be light, 
which would upset the good agreement between the SM 
and the electroweak data.
A large mass splitting in the stop sector 
needs a very artificial fine tuning of the few free parameters in 
the Constrained MSSM, which connects unified masses and couplings
at the GUT scale to their values at the electroweak scale
via RGE, as will be discussed in the next section. 
This is obvious from the mixing matrix in the squark sector 
(see \ref{stopmat}):
if one of the diagonal elements is much larger than $m_t$, 
the off-diagonal
terms of the order  $m_t$ will not cause a mixing and 
the difference between the left- and 
right-handed stops has to come from the evolution of 
the diagonal terms, which depend on 
%
 the Yukawa couplings for top and bottom ($Y_t,Y_b$) 
and the trilinear couplings $A_{t(b)}$. For low $\tan\beta$
$Y_b$ is negligible, while  $A_t$ and $Y_t$ are not free parameters, 
since they  go to 
 fixed point solutions\cite{cmssm}, i.e.
become independent of their values at the GUT scale.
 Therefore there is little freedom to
adjust these parameters within the CMSSM in order to get 
a large splitting between the left- and right-handed stops.  
\normalsize

In addition, problems arise with electroweak symmetry breaking,
since this requires the Higgs mixing parameter $\mu$ 
to be much heavier than the gaugino masses\cite{cmssm}, 
while $R_b$ requires low values of $\mu$ for a 
significant enhancement (since the
chargino has to be preferably Higgsino-like).
In conclusion, within the CMSSM an enhancement of $R_b$ above 
the SM is practically excluded; only if all squark and gaugino 
masses are taken  as free parameters without 
considering the RGE and common values at the GUT scale, then
one can obtain an improvement in $R_b$.

\section{Low energy Constraints in the CMSSM }
\label{constraints}
Within the Constrained Minimal Supersymmetric Model (CMSSM) it is
possible to predict the low energy gauge couplings and masses of the
3~generation particles from the few supergravity inspired 
parameters at the GUT scale. The main ones are $m_0$ and $m_{1/2}$
as discussed in section \ref{mssm}, eq. \ref{free}.
Moreover, the CMSSM predicts electroweak symmetry breaking due to
large radiative corrections from the Yukawa couplings, thus relating
the $Z^0$ boson mass to the top quark mass via the renormalization
group equations (RGE). 
In addition, the cosmological constraints on the 
lifetime of the universe
are considered in the fits.  The new precise measurements of
 the strong coupling constant and the top mass as well as  higher
 order calculations of the $\bsg$ rate 
 exclude perfect fits in the CMSSM,
although  the discrepancies from the best fit 
parameters are below the $2\sigma$ level.

%
 In this analysis  the coupling constants were taken 
from the fits described in the first section.
The new higher order calculations for the important 
$b\rightarrow s\gamma$ rate  indicate that next to 
leading log (QCD) corrections
increase the SM value by about 10\%\cite{misiak}. 
This can be simulated 
in the lowest level calculation by choosing a 
renormalization scale $\mu=0.65m_b$,
which will be done in the following.
 Here we repeat an update of a  previous 
analysis\cite{bek1}  with the new 
input values mentioned above. 
The input data and fitted parameters have been summarized 
in table \ref{t1}.

{\protect\small
\begin{table}[t]
\renewcommand{\rb}[1]{\raisebox{1.35ex}[-1.35ex]{#1}}
\begin{center}
\begin{tabular}{|c||c||c|}
\hline
{input data} & {$\Rightarrow$} &Fit parameters \\
\hline
$\alpha_1,\alpha_2,\alpha_3$ &{ min.}   & \mgut,~\agut  \\
\mt   ~\mb,~$m_\tau$                       & { $\chi^2$} & $Y_t^0,~Y_b^0=Y_\tau^0$ \\           
\mz                          &   & $m_0,~m_{1/2},~\mu$,~ \tb     \\
\bsg                         &   & $A_0$  \\
 $\tau_{universe}$           &   &                            \\
\hline
\end{tabular}
\end{center}
\caption[]{\label{t1}Summary of  input data and fit parameters
for the   global fit from ref. \cite{bek1}. All parameters were fitted simultaneously 
in order to take care of the correlations, but
the GUT scale \mgut~ and corresponding coupling constant \agut~
 are  mainly determined from gauge coupling unification, $\tb$ and
 the Yukawa couplings $Y_{(t,b,\tau)}^0$ at the GUT scale 
 from the masses of the  3th generation, and $\mu$ from electroweak symmetry breaking (EWSB).
 For 
the low $\tb$ scenario the trilinear coupling  $A_0$ is not very relevant, 
but for large $\tb$ it is determined by $\bsg$ and $b\tau$-unification.
The scalar-  and gaugino  masses $(m_0,m_{1/2})$ enter in all observables.}
\end{table}
}
{\protect\normalsize
\begin{table}[ht]
\vspace*{0.27cm}
\renewcommand{\arraystretch}{1.20}
\renewcommand{\rb}[1]{\raisebox{1.75ex}[-1.75ex]{#1}}
\begin{center}
\begin{tabular}{|c|r|r|}
\hline
 \multicolumn{3}{|c|}{ Fitted SUSY parameters and masses in GeV}                       \\
\hline
Symbol & \makebox[2.0cm]{\bf{low $\tb$}} & \makebox[2.5cm]{\bf{high $\tb$}}\\
\hline
\vspace{-0.059cm}
 $m_0$,~   $m_{1/2}$        &  230,~ 225            &  850,~115                 \\
\vspace{-0.059cm}
 $\mu(\mz)$,~$\tan\beta$     &  -880,~1.7           & -190,~30   \\
\vspace{-0.059cm}
  $Y_t(\mt)$,~$A_t(\mz)$    & 0.008,~-370           & 0.006,~86             \\
\hline
\vspace{-0.059cm}
  $\tilde{\chi}^0_1$,~$\tilde{\chi}^0_2$         &  96,~194    & 47,~92      \\ 
\vspace{-0.059cm}
  $ \tilde{\chi}^0_3$,~$ \tilde{\chi}^0_4$      &  509,~519   & 414,~417            \\
\vspace{-0.059cm}
  $\tilde{\chi}^{\pm}_1 $,~$\tilde{\chi}^{\pm}_2$ &  194,~518  & 92,~422      \\
\hline
\vspace{-0.059cm}  
  $\tilde{g}$,~$\tilde{q}$,~$\tilde{l}$   &  558,~545,~563  & 300,~885,~854   \\ 
\hline
\vspace{-0.059cm}
  $       h,~H $                   &  74,~673   & 109,~624          \\
\vspace{-0.059cm}
  $       A,~H^\pm $                   &  680,~684  & 624,~630           \\
\hline
\end{tabular} \end{center}
\caption[]{\label{t2}Values of the fitted SUSY parameters (upper part) and
             corresponding susy masses (lower part) 
             for low and high $\tb$ solutions using the new input data discussed
             in the text.   
             }    
\end{table}
}
\begin{figure}[t]
\vspace{-0.3cm}
    \begin{center}
    \leavevmode
\parbox{0.59\textwidth}{\epsfig{file=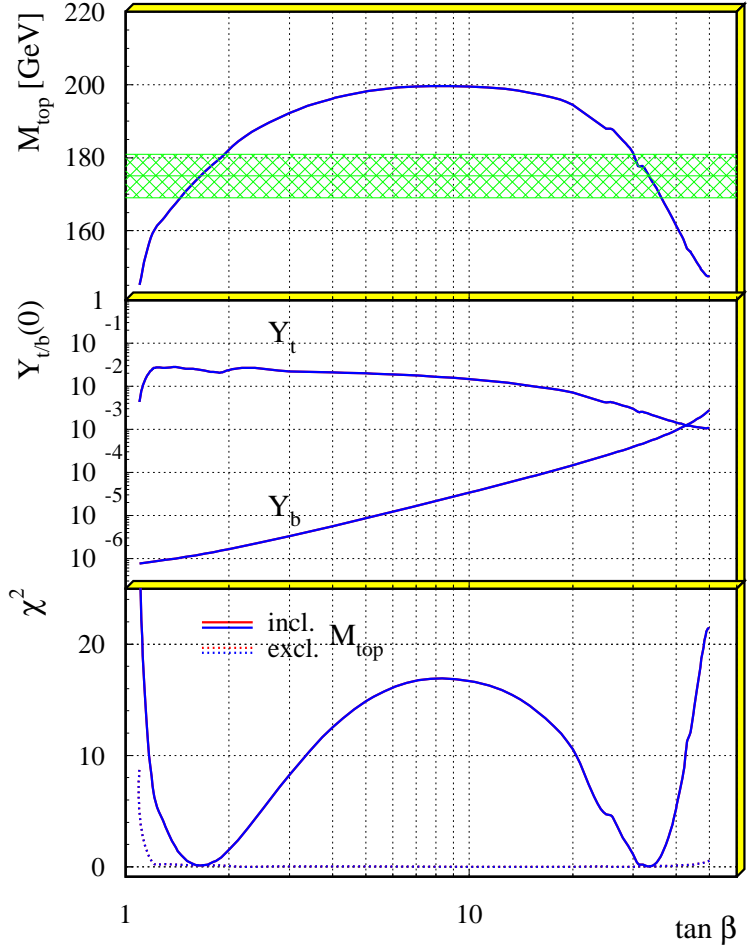,width=0.59\textwidth,
  bbllx=70,bblly=150,bburx=490,bbury=700}   
}
\hfill
\parbox{0.39\textwidth}{
  \caption[]{\label{mttb}The top quark mass as function 
      of $\tb$ (top). 
      The middle part shows the corresponding va\-lues of the Yukawa
      couplings at the GUT scale and the lower part the
      $\chi^2$ values. For $\tb<20$ the Yukawa coup\-ling of the b-quark
     $Y_b$ is small compared to $Y_t$, in which case the top
     quark mass is given by the infrared fixed point solution of $Y_t$.
     For large values of $\tb$ $Y_t$ is reduced by the negative
     corrections from $Y_b$ and $Y_\tau$, which were assumed
     to have common values at the GUT scale ($b-\tau$ unification).  
     If the top constraint ($\mt=175\pm6$, horizontal band) 
      is not applied, all values of $\tb$  are allowed
      (thin dotted lines at the bottom), but if the top mass 
      is constrained to the expe\-rimental value, only the regions
      around $\tb\approx 1.7$ and $\tb\approx 35$ are allowed.
      }
}
 \end{center}
\vspace{-0.5cm}
\end{figure}
%
%
  \begin{figure}[t]
    \vspace{-0.51cm}
\begin{center}
    \leavevmode
\parbox{0.67\textwidth}{
   \epsfig{file=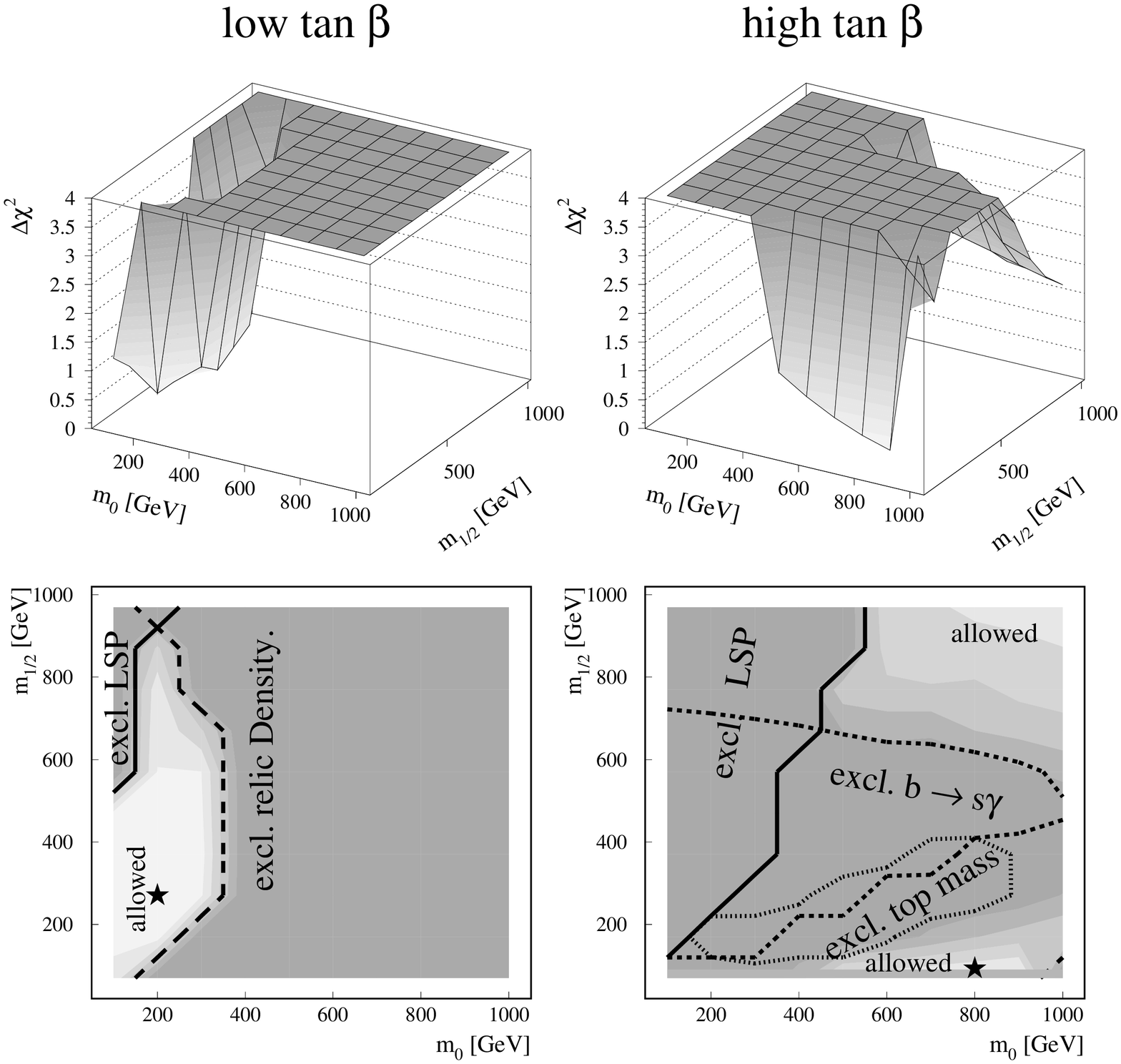,width=0.67\textwidth}
}
\hfill
\parbox{0.3\textwidth}{
\caption[]{\label{\figIII}Contours of the  $\chi^2$-distribution for 
  the low and high $\tb$ solutions in the $m_0$ versus $m_{1/2}$
  plane.  The different shades indicate
  steps of $\Delta\chi^2 = 1$, so basically only the light shaded
  region is allowed.
      The stars indicate the optimum solution. 
      Contours enclose domains excluded by the parti\-cular
      constraints used in the analysis.
}}
\end{center}
\vspace{-1.0cm}
\end{figure}

%
%
%
%
\vspace{0.1cm}
{\bf\underline{Constraints from Gauge 
Coupling Unification}}\vspace{0.2cm}\\
The most  restrictive constraints are
the coupling constant unification and the requirement that the
unification scale has to be above $10^{15}$ GeV from the proton
lifetime limits, assuming decay via s-channel exchange of heavy
gauge bosons. They exclude the SM~\cite{abf} as well
as many other models~\cite{abfI,yana}.
%
\vspace{0.1cm}\\
{\bf\underline{Constraints from the top mass}}\vspace{0.2cm}\\
\label{sec:top}
In the MSSM the top mass is given by: 
\begin{eqnarray}
m_t^2=4\pi Y_t v^2 \frac{\tan^2\beta}{1+\tan^2\beta}.\label{topmass}
\end{eqnarray} 
The top  Yukawa coupling $Y_t$ is given by the RGE, 
which shows a {\it fixed point} behaviour, i.e.
its low energy value is independent of its value 
at the GUT scale\cite{MSSM}, but only
determined by the known {\it gauge} couplings.
Since the VEV of the Higgs field $v=174 $ GeV is known from 
the $Z^0$ mass, all parameters except $\tan\beta$ are known, so 
the MSSM predicts the top (pole) mass to be:
 \begin{eqnarray}
m_t^2\approx (205~{\rm GeV})^2 \sin^2\beta.\label{topmass1}
\end{eqnarray} 
The  maximum possible topmass is around 205 GeV 
and a top mass of 175 GeV corresponds
to $\tan\beta\approx 1.5$, as shown in the top 
part of fig. \ref{\figI}.
For large values of $\tan\beta$ the bottom and 
$\tau$ Yukawa couplings become large
too (see middle part of  fig. \ref{\figI}) and the top  
Yukawa coupling  cannot be predicted
from the gauge couplings alone. However, if one assumes 
$b-\tau$ unification ($Y_b=Y_\tau$ at the GUT scale), 
one  finds a second large $\tan\beta$ 
 solution, as shown in the  top part of fig. \ref{\figI} too, so
for $m_t=175\pm6$ GeV only two regions of $\tb$ give an
acceptable $\chi^2$, as shown in the bottom part of
fig.~\ref{\figI}.
%
%
%
\vspace{0.1cm}\\
{\bf \underline{Electroweak Symmetry Breaking (EWSB)}}\vspace{0.2cm}\\
\label{sec:elwsb}
Radiative corrections can trigger spontaneous symmetry breaking in the
electroweak sector.  In this case the Higgs potential does not have
its mi\-nimum for all fields equal zero, but the 
minimum is obtained for
non-zero vacuum expectation values of the fields.  
Minimization  of the Higgs potential yields: 
\begin{equation}
\label{defmz}
\frac{\mz^2}{2}=\frac{m_1^2+\Sigma _1-(m_2^2+\Sigma _2) \tan^2\beta}
{\tan^2\beta-1},
\end{equation} 
where $m_{1,2}$ are the mass terms in the Higgs potential
and  $\Sigma_1$ and $\Sigma_2$ their radiative corrections.
Note that the radiative corrections are needed, since unification
at the GUT scale with $m_1=m_2$ would lead to $M_Z<0$.
In order to obtain $M_Z>0$ one needs to have 
$m_2^2+\Sigma_2<m_1^2+\Sigma_1$ 
%
%
 which happens at low energy since $\Sigma_2$
($\Sigma_1$) contains large negative corrections proportional to $Y_t$
($Y_b$) and $Y_t \gg Y_b$. 
Electroweak symmetry breaking for the large $\tb$ 
scenario is not so easy,
since  eq. \ref{defmz} can be rewritten as:
\begin{eqnarray} \label{defmz_2}
  \tan^2\beta & = & \frac{m_1^2+\Sigma_1 +
    \frac{1}{2}M_Z^2} {m_2^2 + \Sigma_2 + \frac{1}{2}M_Z^2}.
\end{eqnarray}
For large $\tb$  $Y_t \approx Y_b$, so $\Sigma_1
\approx \Sigma_2$ (see fig. \ref{\figI}). Eq.~\ref{defmz_2} then
requires the starting values of $m_1$ and $m_2$ to be different in
order to obtain a large value of $\tan\beta$, which could happen
if the symmetry group above the GUT scale has a larger rank than
the SM, like e.g. SO(10)\cite{pok1}. 
In this case the quartic interaction
 (D-) terms in the Higgs potential can generate quadratic mass terms,
if the Higgs fields develop non-zero VEVs after spontaneous 
symmetry breaking.

Alternatively, one has to assume the simplest GUT group $SU(5)$, 
which has the same rank as the SM, so no additional 
groups are needed to break SU(5)
and consequently no D-terms are generated. 
In this case EWSB can only be 
generated, if $Y_b$ is sufficiently below $Y_t$, in which case the
different running of $m_1$ and $m_2$
is sufficient  to generate EWSB.
The resulting SUSY mass spectrum is not very sensitive to the
two alternatives for obtaining  $m_1^2 +\Sigma_1 > m_2^2 + \Sigma_2$:
either through a splitting between $m_1$ and $m_2$ 
already at the GUT scale via D-terms
or by generating a difference  via the radiative corrections. 
%
%
\vspace{0.2cm}\\
{\bf\underline{Discussion of the remaining constraints}}\vspace{0.2cm}\\
In fig.~\ref{\figIII} the total $\chi^2$ distribution is shown as a
function of $\mze$ and $\mha$ for the two values of $\tb$ determined
above. One observes  minima at $\mze,\mha$ around (200,270) and
(800,90), as indicated by the stars. These curves were still 
produced with the data from last year. With the new coupling constants
one finds slightly different minima, as given in 
  table~\ref{t2}. In this case the minimum $\chi^2$ is not as good, since the
fit wants $\as\approx 0.125$, i.e. about 1.6$\sigma$ above the measured
LEP value and the calcaluted $\bsg$ rate is above the experimental value too,
if one takes as renormalization scale $\mu\approx 0.65 m_b$. At this
 scale  the next higher order corrections, as calculated by \cite{misiak},
are minimal.  
The contours in fig.~\ref{\figIII} show the regions excluded by
different constraints used in the analysis:
\vspace{0.1cm}\\
 \underline{\bf LSP Constraint:} 
The requirement that the LSP is
  neutral excludes the regions with small $m_0$ and relatively large
  $m_{1/2}$, since in this case one of the scalar staus becomes the LSP
  after mixing via the off-diagonal elements in the mass matrix. 
  The LSP constraint is especially effective at
  the high \tb~ region, since the off-diagonal element 
  in the stau mass matrix 
  is proportional
  to $A_t m_0 - \mu\tan\beta$.
\vspace{0.1cm}\\
 \underline{\bf \bsg Rate:} 
  At low \tb~ the \bsg~ rate is close
  to its SM value for most of the plane. The charginos and/or the
  charged Higgses are only light enough at small values of $m_0$ and
  $m_{1/2}$ to contribute significantly. The trilinear couplings
  were found to play a negligible role for low \tb. 
  However, for large $\tb$ the trilinear coupling needs to be
  left free, since it is difficult to fit simultaneously $\bsg$, $m_b$
  and $m_\tau$. The reason is that the corrections to $m_b$ are
  large for large values of \tb~ due to the large contributions
  from $\tilde{g}-\tilde{q}$ and $\tilde{\chi}^\pm - \tilde{t}$
  loops proportional to $\mu\tb$. They 
  become of the order of 10-20\%. In
  order to obtain $m_b(M_Z)$ as low as 2.84 GeV, these corrections
  have to be negative, thus requiring $\mu$ to be negative.
  The \bsg rate is too large in
  most of the parameter region for large \tb, because of the
dominant chargino contribution, which is proportional to $A_t\mu$.
For positive (negative) values of $A_t\mu$ this leads to a larger
(smaller) branching ratio \Bbsg than  for the Standard Model with two
Higgs doublets.
In order to reduce
  this rate one needs $A_t(M_Z)>0$ for $\mu<0$.
  Since for large \tb~ $A_t$ does not show a fixed point behaviour,
  this is possible.

\vspace{0.1cm}
 \underline{\bf Relic Density:}
  The long lifetime of the universe requires a  mass density below the
  critical density, else the overclosed universe would have collapsed long ago.
  This requires that the contribution from the LSP to the relic density has to be 
  below the critical density, which can be achived if the annihilation rate
  is high enough. Annihilation into electron-positron pairs proceeds either
  through t-channel selectron exchange or through s-channel $Z^0$ exchange
  with a strength given by the Higgsino component of the lightest neutralino.
  For the low \tb~ scenario the value of
  $\mu$ from EWSB is large\cite{bek1}. In this case
  there is little mixing between the higgsino- and gaugino-type neutralinos as is
  apparent from the neutralino mass matrix: for $|\mu| \gg M_1 \approx
  0.4 m_{1/2}$ the mass of the LSP is simply $0.4 m_{1/2}$ and the
  ``bino'' purity is 99\%\cite{bek1}. For the high \tb~
  scenario $\mu$ is much smaller  and the
  Higgsino admixture becomes larger.  This leads to an enhancement of
  $\tilde\chi^0-\tilde\chi^0$ annihilation via the s-channel Z boson
  exchange, thus reducing the relic density.  As a result, in the
  large $\tb$ case the constraint $\Omega h_0^2 < 1$ is almost always
  satisfied unlike in the case of low $\tb$.

\begin{figure}[t]
 \begin{center}
\parbox{0.6\textwidth}{
  \leavevmode
  \epsfig{file=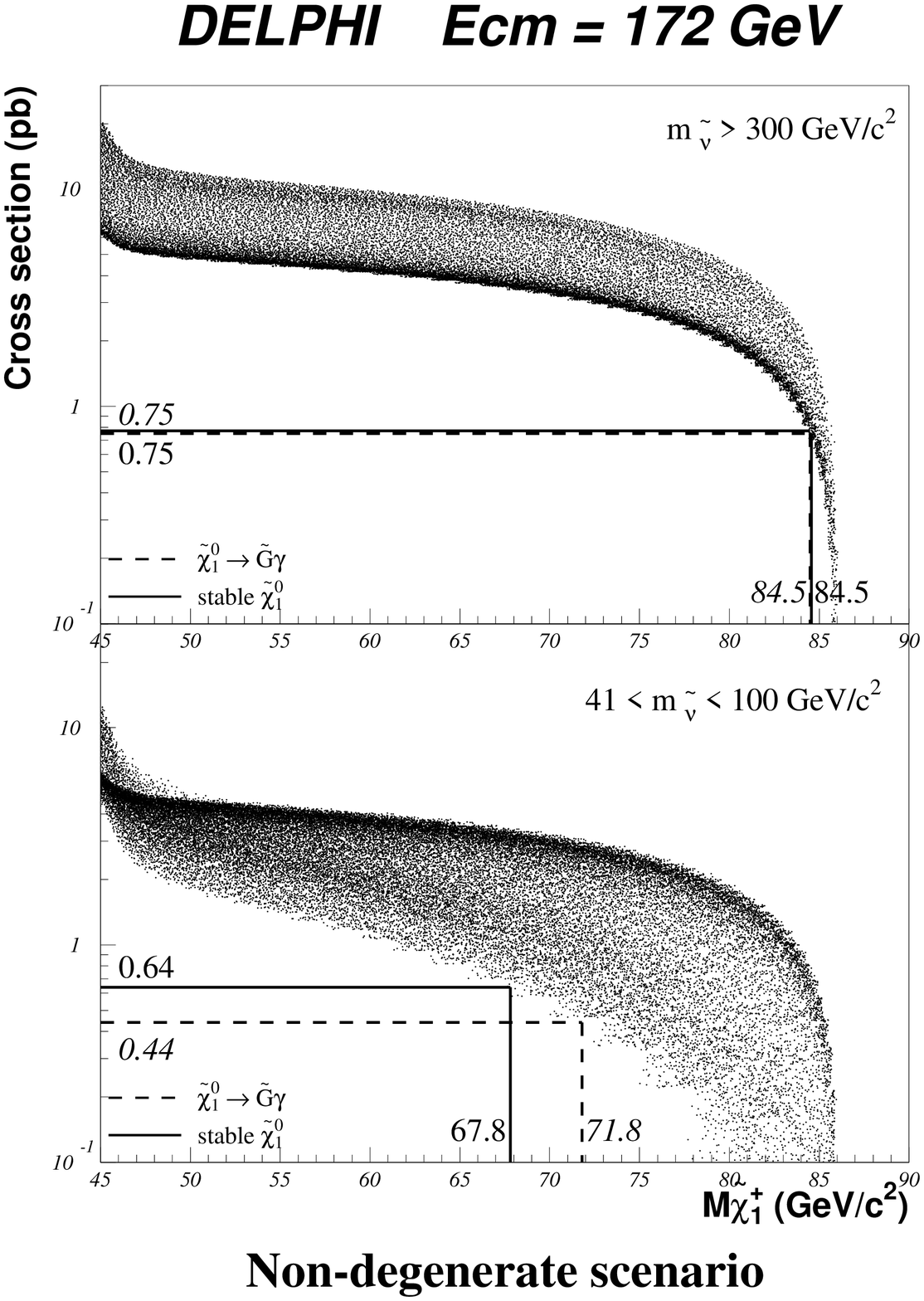,width=0.6\textwidth}
}
\hfill
\parbox{0.39\textwidth}{
\caption[]{\label{chargino_delphi}
Chargino limits from DELPHI for 4 different cases:
for heavy sneutrinos with stable  and unstable 
neutralinos the chargino mass
is above  84.5 GeV at 95\% C.L. (upper part); 
As indicated, the unstable neutralino is assumed to decay
into a photon and gravitino.
For light sneutrinos
the negative interference between s- and t-channel reduces the cross
section, thus leading  to worse limits as shown in the bottom part.
It is assumed that the lightest chargino is non-degenerate 
with the LSP.
In case the  lightest chargino is Higgsino-like, implying
$\mu<M_2$ (see eq. \ref{chamat}), the chargino can be degenerate
with the LSP, but the  t-channel sneutrino  is suppressed in this case.
If the degeneracy is less than 5 GeV, limits rather close to the
kinematic limit are obtained.
From \cite{delphi}.}
}
 \end{center}
\vspace{-0.6cm}
\end{figure}
\begin{figure}[t]
\vspace{-1.7cm}
 \begin{center}
\parbox{0.62\textwidth}{
  \leavevmode
  \epsfig{file=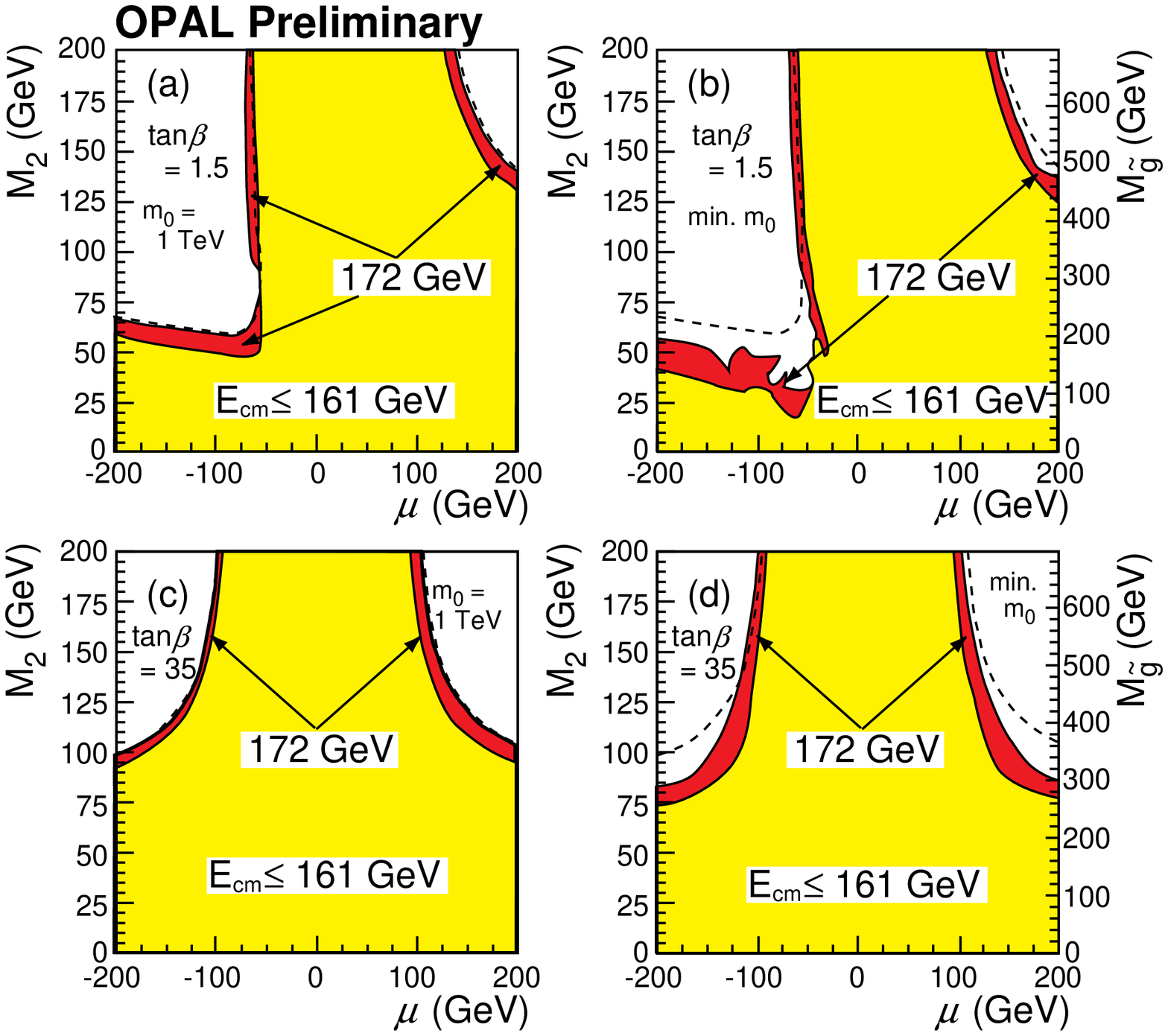,width=0.62\textwidth 
  }
}
\hfill
\parbox{0.37\textwidth}{
\caption[]{\label{mum2_opal}
$M_2$ versus $\mu$ from OPAL for LEP data from $\sqrt{s}=161$ and 172 GeV.
Note that for chargino/neutralino searches the reach in 
parameter space increases only li\-near with
energy in contrast to the Higgs searches. From \cite{opal}.}
}
 \end{center}
%
\vspace{0.3cm}
 \begin{center}
\parbox{0.62\textwidth}{
  \leavevmode
  \epsfig{file=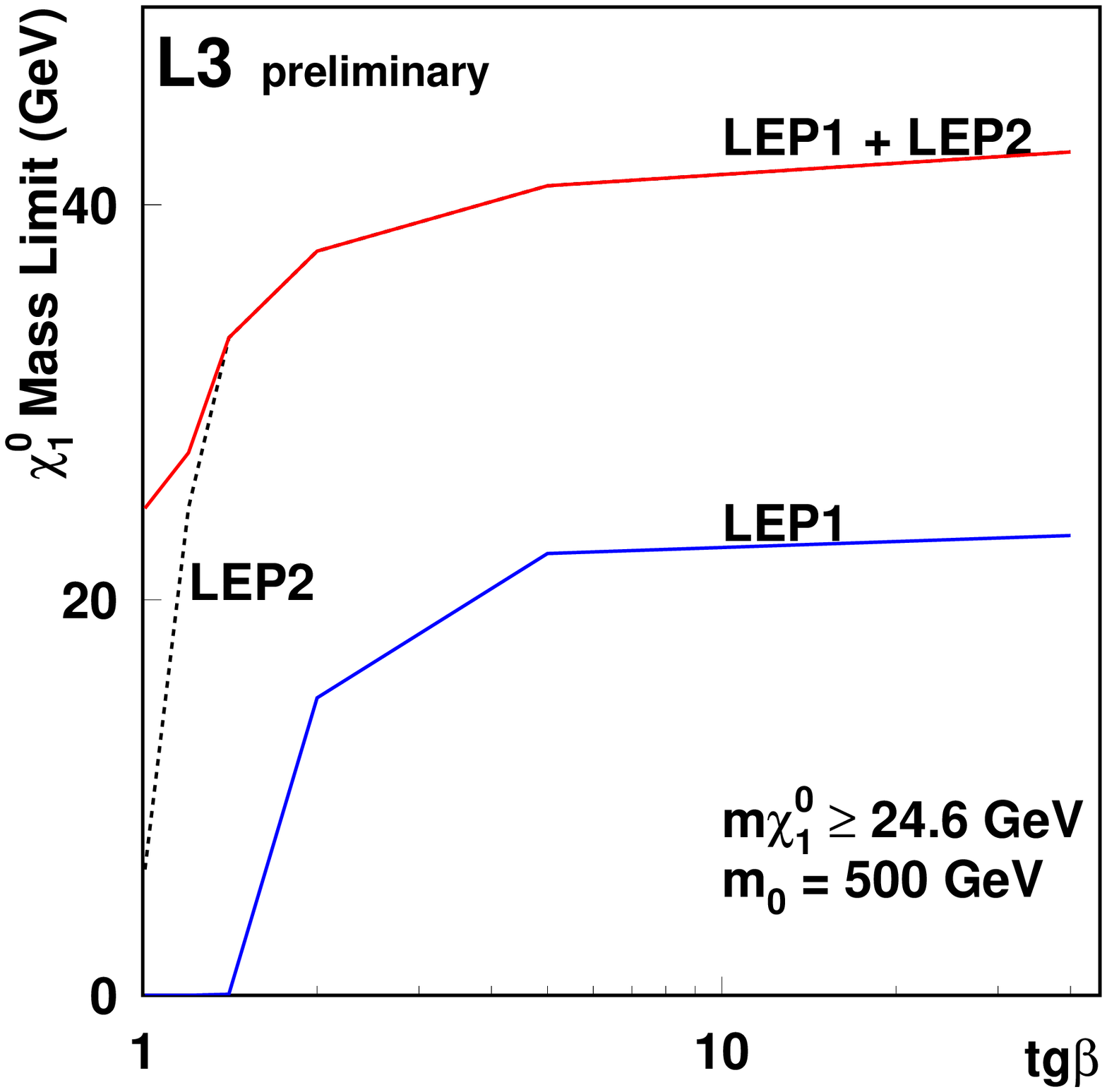,width=0.58\textwidth}
}
\parbox{0.37\textwidth}{
\caption[]{\label{neutralino_l3}
The mass of the lightest stable neutralino, assumed to be the  LSP,
 is related to the lightest 
chargino mass via the RGE equations, which connect common gaugino  
masses at the GUT scale to the electroweak scale. Combining the
search limits of chargino and neutralino pair production leads to 
a lower mass limit of 24.6 GeV for the invisible LSP
for all values of $\tan\beta$, provided the lightest 
sneutrino is heavy.
For light sneutrinos the negative t- and s-channel interference
reduce the chargino cross section, thus reducing the LSP limit. 
From \cite{l3}.
}
}
 \end{center}
\vspace{-0.6cm}
\end{figure}
\begin{figure}[t]
 \begin{center}
   \vspace{-2.5cm}
\parbox{0.7\textwidth}{
  \leavevmode
  \epsfig{file=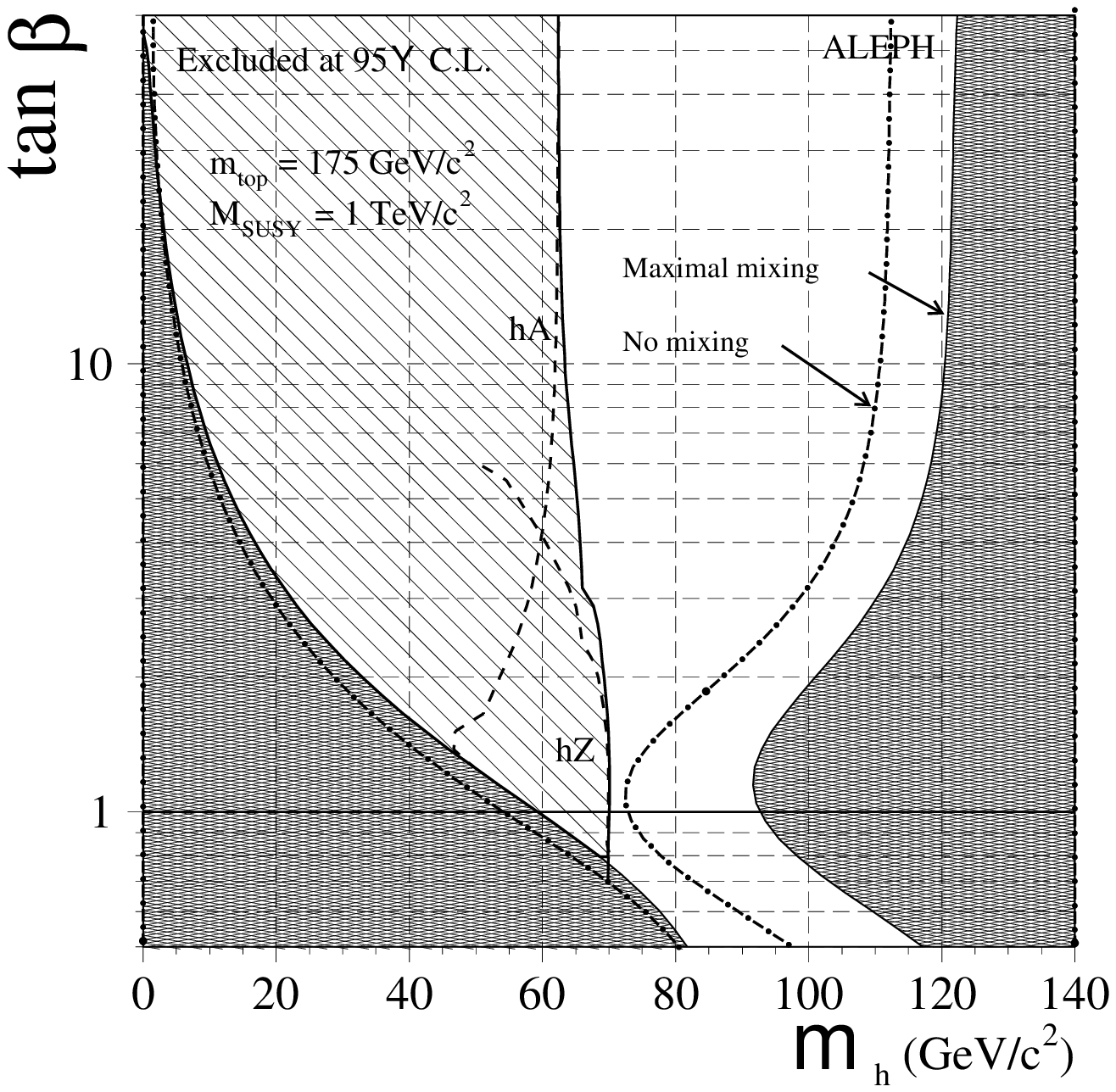,width=0.7\textwidth }
}
\parbox{\textwidth}{
\caption[]{f\label{higgs_aleph}$\tan\beta$ versus
the Higgs mass  from ALEPH\cite{aleph}.
 The dashed area is excluded by the search 
for the $hZ$ and $hA$ final
states, which require both $m_h$ and $m_A$ to be above 
62.5 GeV at 95\% C.L. The dark regions are excluded in case of 
large mixing in the stop sector, the solid line in case of no mixing.
In the constrained MSSM the mixing is usually small, so for small
$\tan\beta$ the combined data from all LEP experiments will 
exclude the low $\tan\beta$ scenario. The region for $2<\tan\beta<40$
is excluded from the solution of the RGE for 
the top Yukawa coupling, as 
shown in fig. \ref{\figI}.
}
}
 \end{center}
%
%
%
%
%
  \begin{center}
    \leavevmode
    \epsfxsize=13cm
    \epsffile{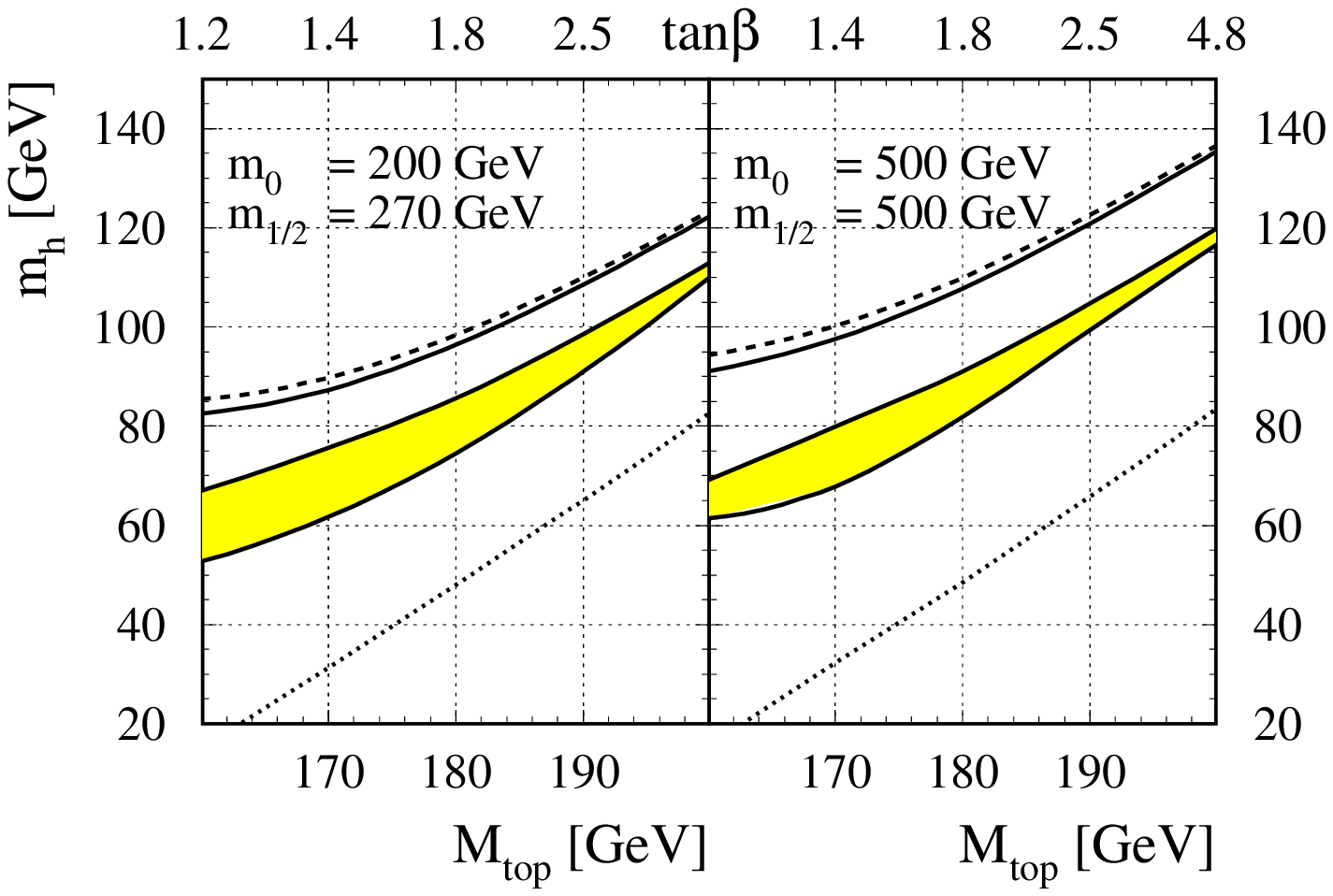}
  \end{center}
  \caption[]{\label{f4}The mass of the lightest CP-even Higgs as 
    function of the top mass at Born level (dotted lines), 
    including complete one-loop contributions of all particles
    (dashed lines). Two-loop contributions reduce the one-loop
    corrections significantly as shown by the dashed area
    (the upper boundary corresponds to $\mu>0$, the lower one
    to $\mu<0$). The solid line just
    below the dashed line is the one-loop prediction 
    from the third generation only,
    which apparently gives the main contribution. The upper scale
    indicates the value of $\tb$, as calculated from the top mass
    (eq. \ref{topmass1}).
    }
\end{figure}
%

\section{Discovery Potential  at LEP II}\label{lep2}
All LEP experiments have been searching for 
the sparticles and Higgs bosons predicted by the MSSM.
Table \ref{t2} shows that charginos, neutralinos and the lightest
Higgs belong to the lightest particles in the MSSM, so we 
will concentrate on these searches and show only a few
typical results for each experiment keeping
in mind that the other experiments have usually similar
results on the same channel.

Charginos are expected to  be easy to discover, 
since they will be pair produced
with a large cross section of several $pb$ and
lead to events with characteristic decays similar to $W^\pm$ pairs
plus missing energy.
The typical limits are close to the beam limit, as shown in Fig.
\ref{chargino_delphi}
by recent results from DELPHI\cite{delphi}.
Since the chargino mass depends on the SUSY parameters 
$\mu, M_2$ and $\tb$ these limits can be shown
as contours in the $\mu-M_2$ plane for a given value of $\tb$,
as shown in Fig. \ref{mum2_opal} for LEP data at 161 and 172 GeV 
centre of mass energies from OPAL\cite{opal}.
If one assumes the GUT relation $M_2\approx 2M_1$ (eq. \ref{gaugino})
 the neutralino
limits are related to the chargino limits.
Combining it with direct neutralino searches, both at LEP I and LEP II,
L3 finds a lower limit on the neutralino mass of 24.6 GeV\cite{l3}
as shown in Fig. \ref{neutralino_l3}.
The Higgs mass is a function of the pseudoscalar Higgs mass $m_A$,
$\tan\beta$ and the topmass via the radiative corrections.
Higgs bosons can be produced through Higgs-strahlung $e^+e^-\rightarrow hZ$
and associated production $e^+e^-\rightarrow hA$. The first one 
is proportional to $\sin^2(\beta-\alpha)$, while the second
one to $\cos^2(\beta-\alpha)$, so the total cross section is
independent of the mixing angles $\beta-\alpha$. 
If one searches for both processes one can find a Higgs limit
independent of $\tan\beta$, as shown for the ALEPH data in fig. 
\ref{higgs_aleph} (from ref. \cite{aleph}).

The Higgs mass  depends on the top mass as shown in fig. \ref{f4}.
Here the most significant second order corrections to the Higgs mass
have been incorporated \cite{ll}, which reduces the Higgs mass by
about 15 GeV \cite{bekhiggs}. In this case 
the Higgsmass is below 90 GeV, provided the top mass is below 
180 GeV (see fig. \ref{f4}), which implies that 
the foreseen LEP energy of
192 GeV is sufficient to cover the whole parameter space.
\subsection{Summary}
In summary,     
in the Constrained Minimal Supersymmetric Model (CMSSM) the allowed 
region of the GUT scale parameters and the corresponding SUSY mass
spectra for the low and high $\tb$ scenario have been determined from
a combined fit to the low energy data on couplings, quark and lepton
masses of the third generation, the electroweak 
scale $\mz$, $\bsg$, and
the lifetime of the universe.
The new precise determinations of the strong coupling constant 
$\as=0.120\pm0.003$ are slightly below the preferred CMSSM 
fit value of about 0.125. In addition,  the observed $\bsg$ value
of $(2.32\pm0.6)10^{-4}$ is below the predicted   value, 
at least for the SM  $(3.2\cdot10^{-4})$
and the low $\tb$ scenario of the MSSM. 

The lightest particles  preferred by these fits 
are charginos and higgses.
The charginos are preferably light in case of the high $\tb$ scenario,
while the lightest higgs wiil be within reach of LEP II
in case of the low $\tb$ scenario (see fig. \ref{mttb}).
So the light $\tb$ scenario of the CMSSM can be confirmed
or excluded at LEP II (provided the top mass is indeed below 180 GeV),
while the complete parameter space for the high $\tb$ scenario
will become only accesible at future accelerators.

It should be noted that recent speculation about evidence 
for SUSY from the $ee\gamma\gamma$ event 
observed by  the CDF collaboration\cite{kane}, 
the too high value of $R_b$\cite{EWWG,hollik1} and the 
ALEPH 4-jet events\cite{aleph4jets} has not been confirmed sofar:
\begin{itemize}
\item if the single CDF $ee\gamma\gamma+E_{miss}$ 
event would originate from 
selectron pair production
with the two gammas coming from neutralino decay into
either the LSP or gravitino, one would expect anomalous inclusive
$p\overline{p}\rightarrow \gamma\gamma+E_{miss}+X$ 
production, which has not been observed\cite{carithers}.
\item The $R_b$ anomaly is reduced to ``a-less-than-$2\sigma$-effect''
\cite{EWWG}. 
\item The anomalous ALEPH 4-jet events  have not  
 been confirmed by the other three LEP Collaborations\cite{schlatter}.
\end{itemize}

\vspace{0.2cm}
{\bf\underline{Acknowledgements}}
I want to thank my colleagues from the LEP groups for  helpful discussions and/or
making available data prior to publication, especially
Glen Cowan,  Ralf Ehret, Dmitri Kazakov, Michael Kobel, Sachio Komamyia, Marco Pieri, 
Silvie Rosier, Michael Schmitt, and 
Ulrich Schwickerath.

\clearpage

%
\end{document}